\documentclass[twocolumn,showpacs,preprintnumbers,amssymb,nofootinbib]{revtex4}

\usepackage{graphicx}
\usepackage{bm} 

\begin{document}


\title{Pair of accelerated black holes in anti-de Sitter
background: the AdS C-metric}

\author{\'Oscar J. C. Dias}
\email{oscar@fisica.ist.utl.pt}
\author{Jos\'e P. S. Lemos}
\email{lemos@kelvin.ist.utl.pt}
\affiliation{
Centro Multidisciplinar de Astrof\'{\i}sica - CENTRA,
Departamento de F\'{\i}sica, Instituto Superior T\'ecnico,
Av. Rovisco Pais 1, 1049-001 Lisboa, Portugal
}%

\date{\today}

\begin{abstract}
The anti-de Sitter C-metric (AdS C-metric) is characterized by a
quite interesting new feature when compared with the C-metric in
flat or de Sitter backgrounds. Indeed, contrarily to what happens
in these two last exact solutions, the AdS C-metric  only
describes a pair of accelerated black holes if the acceleration
parameter satisfies $A> 1/ \ell$, where $\ell$ is the cosmological
length. The two black holes cannot interact gravitationally and
their acceleration is totally provided by the pressure exerted by
a strut that pushes the black holes apart. Our analysis is based
on the study of the causal structure, on the description of the
solution in the AdS 4-hyperboloid in a 5D Minkowski spacetime, and
on the physics of the strut. We also analyze the cases $A= 1/
\ell$ and $A< 1/ \ell$ that represent a single accelerated black
hole in the AdS background.
\end{abstract}


\maketitle
\section{Introduction}

The original C-metric has been found by Levi-Civita in his studies
between 1917 and 1919. During the following decades, many authors
have rediscovered it and studied its mathematical properties (see
\cite{Kramer} for references). In 1963 Ehlers and Kundt
\cite{EhlersKundt} have classified degenerated static vacuum
fields and put this Levi-Civita solution into the C slot of the
table they constructed. From then onwards this solution has been
called C-metric. This spacetime is stationary, axially symmetric,
Petrov type D, and is an exact solution which includes a radiative
term. Although the C-metric had been studied from a mathematical
point of view along the years, its physical interpretation
remained unknown until 1970 when Kinnersley and Walker \cite{KW},
in a pathbreaking work, have shown that the solution describes two
uniformly accelerated black holes in opposite directions. Indeed,
they noticed  that the original solution was geodesically
incomplete, and by defining new suitable coordinates they have
analytically extended it and studied its causal structure. The
solution has a conical singularity in one of its angular poles
that was interpreted by them as due to the presence of a strut in
between pushing the black holes away, or as two strings from
infinity pulling in each one of the black holes. The strut or the
strings lie along the symmetry axis and cause the acceleration of
the black hole pair. This work also included for the first time
the charged version of the C-metric. In an important development,
Ernst  in 1976 \cite{Ernst}, trough the employment of an
appropriate transformation, has removed all the conical
singularities of the charged C-metric by appending an external
electromagnetic field. In this new exact Ernst solution the
acceleration of the pair of oppositely charged  black holes is
provided by the Lorentz force associated to the external field.
The geometrical properties of the C-metric were further
investigated by Farhoosh and Zimmerman \cite{FarhZimm1}, and the
asymptotic properties of the C-metric were analyzed by Ashtekar
and Dray \cite{AshtDray} who have shown that  null infinity admits
a conformal completion, has a spherical section, and moreover that
the causal diagrams drawn in \cite{KW} were not quite accurate.
The issue of physical interpretation of the C-metric has been
recovered by Bonnor \cite{Bonnor1}, but now following a different
approach. He transformed the C-metric into the Weyl form in which
the solution represents a finite line source (that corresponds to
the horizon of the black hole), a semi-infinite line mass
(corresponding to a horizon associated with uniform accelerated
motion) and a strut keeping the line sources apart. By applying a
further transformation that enlarges this solution, Bonnor
confirmed the physical interpretation given in \cite{KW}. Bonnor's
procedure has been simplified by Cornish and Uttley and extended
to include the massive charged solution \cite{CornUtt1}. More
recently, Yongcheng \cite{Yong}, starting from the metric of two
superposed Schwarzschild black holes, has derived the C-metric
under appropriate conditions. The black hole uniqueness theorem
for the C-metric has been proven by Wells \cite{Wells} and the
geodesic structure of the C-metric has been studied by Pravda and
Pravdova \cite{PravPrav2}. The limit when the acceleration goes to
infinity has been analyzed by Podolsk\' y and Griffiths
\cite{PodGrif1} who have shown that in this limit the solution is
analogous to the one which describes a spherical impulsive
gravitational wave generated by a snapping string. We note that
the C-metric is an important and explicit example of a general
class of asymptotically flat radiative spacetimes with
boost-rotation symmetry and with hypersurface orthogonal axial and
boost Killing vectors. The geometric properties of this general
class of spacetimes have been investigated by Bi\v c\' ak and
Schmidt \cite{BicSchm} and the radiative features were analyzed by
Bi\v c\' ak \cite{Bic} (see the recent review of Pravda and
Pravdova  \cite{PravPrav} on this class of spacetimes and the role
of the C-metric).

Relevant generalizations to the C-metric were made by Pleba\'nski
and Demia\'nski in 1976 \cite{PlebDem} and by Dowker, Gauntlett,
Kastor and Traschen  in 1994 \cite{DGKT}. Pleba\'nski and
Demia\'nski, in addition to the mass ($m$) and electromagnetic
charge ($q$), have  included into the solution a NUT parameter, a
rotation and a cosmological constant term ($\Lambda$), and Dowker
et al have further included a dilaton field non-minimally coupled.
Thus, the most general C-metric has eight parameters, so far,
namely, acceleration, mass, electric and magnetic charges, NUT
parameter, rotation, cosmological constant and dilaton field. The
C-metric with mass and electromagnetic charges alone have been
extensively studied as shown above, and from now on we will refer
to it as the flat C-metric (i.e., C-metric with $\Lambda=0$). The
C-metric with a NUT parameter has not been studied, as far as we
know. The flat spinning C-metric has been studied by Farhoosh and
Zimmerman \cite{FarhZimm4}, Letelier and Oliveira \cite{LetOliv}
and by Bi\v c\' ak and Pravda \cite{BicPrav}. In particular, in
\cite{BicPrav} the flat spinning C-metric has been transformed
into the Weyl form and interpreted as two uniformly accelerated
spinning black holes connected by a strut. This solution
constitutes an example of a spacetime with an axial and boost
Killing vectors which are not hypersurface orthogonal. Dowker et
al generalized the flat C-metric and  flat Ernst solution to
include a dilaton field and applied these solutions for the first
time in the context of quantum pair creation of black holes, that
once created accelerate apart.

In what concerns the cosmological C-metric introduced in
\cite{PlebDem}, the de Sitter (dS) case ($\Lambda>0$) has been
analyzed by Podolsk\' y and Griffiths \cite{PodGrif2}, whereas the
anti-de Sitter (AdS) case ($\Lambda<0$)  has been studied, in
special instances, by Emparan, Horowitz and Myers \cite{EHM1} and
by Podolsk\' y \cite{Pod}. In general the C-metric (either flat,
dS or AdS) describes a pair of accelerated black holes. Indeed, in
the flat and dS backgrounds this is always the case. However, in
an AdS background the situation is not so simple and depends on
the relation between the acceleration $A$ of the black holes and
the cosmological length $\ell$. Since the AdS C-metric presents
such peculiar features it deserves a careful analysis. It is our
intention in this paper to fully study, in its most general form,
the AdS C-metric with mass, charge and cosmological constant. One
can divide the study into three cases, namely, $A<1/\ell$,
$A=1/\ell$ and $A>1/\ell$. The $A<1/\ell$ case was the one
analyzed by Podolsk\' y \cite{Pod}, and the $A =1/\ell$ case has
been investigated by Emparan, Horowitz and Myers \cite{EHM1},
which has acquired an important role since the authors have shown
that, in the context of a lower dimensional Randall-Sundrum model,
it describes the final state of gravitational collapse on the
brane-world. The geodesic structure of this solution has been
studied by Chamblin \cite{Chamb}. Both cases, $A < 1/\ell$ and $A=
1/\ell$, represent one  single accelerated black hole.  The case
$A> 1/\ell$ has not been fully studied and its physical
interpretation is not yet firmly established, although it has been
applied, in addition to the flat and dS cases, in pair creation of
black holes by Hawking, Horowitz and Ross \cite{HHR}, and by Mann
\cite{MannAdS} (see \cite{Osc} for a review). The purpose of this
article is to establish that the $A> 1/\ell$ AdS C-metric
describes a pair of accelerated black holes in an AdS background.
This aim will be achieved through a thorough analysis of the
causal structure of the solution, together with the description of
the solution in the AdS 4-hyperboloid, and the study of the
strut's physics.

The plan of this article is as follows. In section \ref{sec:Int}
we present the AdS C-metric and analyze its curvature and conical
singularities. In section \ref{sec:PD} we study the causal
diagrams of the solution. In section \ref{sec:Phys_Interp} we give
and justify a physical interpretation to the solution. The
description of the solution in the AdS 4-hyperboloid and the
physics of the strut are analyzed. These two sections,
\ref{sec:PD} and \ref{sec:Phys_Interp}, are highly related and so,
in order to fully understand each of them, the reading of the
other is required. Finally, in section \ref{sec:Conc} concluding
remarks are presented.

\section{\label{sec:Int}GENERAL PROPERTIES OF THE A\lowercase{d}S C-METRIC}

\subsection{The AdS C-metric}
The AdS C-metric, i.e., the C-metric with negative cosmological
constant $\Lambda$, has been obtained by Pleba\'nski and
Demia\'nski \cite{PlebDem}. For zero rotation and zero NUT
parameter it is given, according to \cite{PlebDem} (see also
\cite{MannAdS}), by
\begin{equation}
 d s^2 = 1/(\tilde{x}+\tilde{y})^2 (-{\cal F}d\tilde{t}^2+
 {\cal F}^{-1}d\tilde{y}^2+{\cal G}^{-1}d\tilde{x}^2+
 {\cal G}d\tilde{z}^2)\:,
 \label{C-PD}
 \end{equation}
 where
 \begin{eqnarray}
 & &{\cal F}(\tilde{y}) = -\Lambda/6-\tilde{A}^2+ \tilde{y}^2
              -2m\tilde{y}^3+q^2\tilde{y}^4, \nonumber\\
 & &{\cal G}(\tilde{x}) = -\Lambda/6+\tilde{A}^2
 -\tilde{x}^2-2m\tilde{x}^3-q^2\tilde{x}^4.
 \label{FG-1}
 \end{eqnarray}
The meaning of parameters $\tilde{A}$, $m$, and $q$ will be
clarified soon. For the benefit of comparison with the flat
C-metric, we note that when $\Lambda$ vanishes we have ${\cal
F}(\tilde{y})=-{\cal G}(-\tilde{y})$. It is now convenient to
redefine the parameter $\tilde{A}$ as
$-\Lambda/6+\tilde{A}^2\equiv A^2$, together with the coordinate
transformations: $\tilde{t}=t/A, \tilde{y}=Ay, \tilde{x}=Ax$ and
$\tilde{z}=z/A$. With these redefinitions, the gravitational field
of the AdS C-metric is written as
\begin{equation}
 d s^2 = [A(x+y)]^{-2} (-{\cal F}dt^2+
 {\cal F}^{-1}dy^2+{\cal G}^{-1}dx^2+
 {\cal G}dz^2)\:,
 \label{C-metric}
 \end{equation}
 where
 \begin{eqnarray}
 & &{\cal F}(y) = {\biggl (}\frac{1}{\ell^2A^2}-1{\biggl )}
                     +y^2-2mAy^3+q^2A^2y^4, \nonumber\\
 & &{\cal G}(x) = 1-x^2-2mAx^3-q^2 A^2 x^4\:,
 \label{FG}
 \end{eqnarray}
and the non-zero components of the electromagnetic vector
potential, $A_{\mu}dx^{\mu}$, are given by
\begin{eqnarray}
 A_{\rm t}=-e \,y \:, \;\;\;\;\;\;\;A_{\rm z}=g \, x  \:.
\label{potential}
\end{eqnarray}
 This solution depends on four parameters namely, $A$ which
 is the acceleration of the black hole,  $m$ which is
 interpreted  as the ADM mass  of the
 non-accelerated black hole, $q$ which is
 interpreted as the ADM electromagnetic charge of the
 non-accelerated black hole and, in general, $q^2=e^2+g^2$ with $e$ and $g$
 being the electric and magnetic charges, respectively,  and finally the cosmological
 length $\ell^2\equiv3/|\Lambda|$. The meaning attributed to
 the parameter $A$ will be understood in section \ref{sec:Phys_Interp}, while the
 physical interpretation given to the parameters $m$ and $q$ is
 justified in the Appendix. We will consider the case $A>0$.

 The coordinates used in Eqs. (\ref{C-metric})-(\ref{potential})
 to describe the AdS C-metric are useful to understand the geometrical
properties of the spacetime, but they hide the physical
interpretation of the solution. In order to understand the
physical properties of the source and gravitational field we will
introduce progressively new coordinates more suitable to this
propose, following the approach of Kinnersley and Walker \cite{KW}
and Ashtekar and Dray \cite{AshtDray}. Although the alternative
 approach of Bonnor simplifies in a way the interpretation, we
cannot use it were since the cosmological constant prevents such a
coordinate transformation into the Weyl form.

\subsection{\label{sec:CurvSing}Radial Coordinate. Curvature Singularities}

We start by defining a
coordinate $r$ as
\begin{equation}
 r = [A(x+y)]^{-1} \:.
 \label{r}
 \end{equation}
In order to interpret this coordinate as being a radial
coordinate, we calculate a curvature invariant of the metric,
namely the Kretschmann scalar,
\begin{eqnarray}
        R_{\mu\nu\alpha\beta}R^{\mu\nu\alpha\beta} &=&
        \frac{24}{\ell^2}
        +\frac{8}{r^8}{\biggl [}6m^2r^2+12m q^2(2Axr-1)r
                                 \nonumber \\
        & &
        +q^4(7-24Axr+24A^2x^2r^2){\biggr ]}
          \:.
                             \label{R2}
\end{eqnarray}
Clearly, this curvature invariant is equal to $24/\ell^2$ when the
mass $m$ and charge $q$ are both zero. When at least one of these
parameters is not zero, the curvature invariant diverges at $r=0$,
revealing the presence of a curvature singularity. Moreover, when
we take the limit $r\rightarrow \infty$, the curvature singularity
approaches the expected value for a spacetime which is
asymptotically AdS. Therefore it is justified that $r$ is
interpreted as a radial coordinate.

\subsection{\label{sec:ConSing}Angular Surfaces. Conical Singularities}

To gain more insight into the physical nature of the AdS C-metric
we now turn our attention into the angular surfaces $t=$constant
and $r=$constant, onwards labelled by $\Sigma$. In this section we
follow \cite{KW}. In order to have the AdS C-metric with correct
signature, $(-+++)$, one must restrict the coordinate $x$ to a
domain on which the function ${\cal{G}}(x)$ is non-negative [see
Eq. (\ref{C-metric})]. The shape of this function depends
crucially on the three parameters $A$, $m$, and $q$. In this work
we will select only the ranges of these three parameters for which
${\cal{G}}(x)$ has at least two real roots, $x_\mathrm{s}$ and
$x_\mathrm{n}$ (say), and demand $x$ to belong to the range
$[x_\mathrm{s},x_\mathrm{n}]$ where ${\cal{G}}(x)\geq 0$. This
restriction has the important advantage of allowing us to endow
the angular surfaces $\Sigma$ with the topology of a compact
surface. In these surfaces we now define two new coordinates,
\begin{eqnarray}
 \theta &=& \int_{x}^{x_\mathrm{n}}{\cal{G}}^{-1/2}dx \:,
                                 \nonumber \\
 \phi &=& z/\kappa \:,
                             \label{ang}
\end{eqnarray}
where $\phi$ ranges between $[0,2\pi]$ and $\kappa$ is an
arbitrary positive constant which will be needed later when
regularity conditions at the poles are discussed. The coordinate
$\theta$ ranges between the north pole,
$\theta=\theta_\mathrm{n}=0$, and the south pole,
$\theta=\theta_\mathrm{s}$ (not necessarily at $\pi$). With these
transformations the metric restricted to the surfaces $\Sigma$, $d
\sigma^2 = r^2 ({\cal G}^{-1}dx^2+{\cal G}dz^2)$, takes the form
\begin{equation}
 d \sigma^2
 = r^2{\bigl (} d\theta^2 + \kappa^2{\cal{G}}d\phi^2{\bigr )}\:.
 \label{ang-metric}
 \end{equation}
When $A=0$ or when both $m=0$ and $q=0$, Eq. (\ref{ang}) gives
$x=\cos{\theta}$, ${\cal G}=1-x^2=\sin^2{\theta}$ and if we use
the freedom to put $\kappa \equiv 1$, the metric restricted to
$\Sigma$ is given by $d\sigma^2= r^2 ( d\theta^2 +
\sin^2{\theta}\,d\phi^2 )$. This implies that in this case the
angular surface is a sphere and justifies the label given to the
new angular coordinates defined in Eq. (\ref{ang}). In this case
the north pole is at $\theta_\mathrm{n}=0$ or $x_\mathrm{n}=+1$
and the south pole is at $\theta_\mathrm{s}=\pi$ or
$x_\mathrm{s}=-1$. In the other cases $x$ and $\sqrt{\cal G}$ can
always be expressed as elliptic functions of $\theta$. The
explicit form of these functions is of no need in this work. All
we need to know is that these functions have a period given by
$2\theta_\mathrm{s}$.

As we shall see, the regularity analysis of the metric in the
region $[0,\theta_\mathrm{s}]$ will play an essential role in the
physical interpretation of the AdS C-metric. The function ${\cal
G}$ is positive and bounded in $]0,\theta_\mathrm{s}[$ and thus,
the metric is regular in this region between the poles. We must be
more careful with the regularity analysis at the poles, i.e., at
the roots of ${\cal G}$. Indeed, if we draw a small circle around
the north pole, in general, as the radius goes to zero, the limit
circunference/radius is not $2\pi$. Therefore, in order to avoid a
conical singularity at the north pole one must require that
$\delta_\mathrm{n}=0$, where
\begin{equation}
 \delta_\mathrm{n} \equiv 2\pi {\biggl (}1-\lim_{\theta \rightarrow 0}
 \frac{1}{\theta}\sqrt{\frac{g_{\phi\phi}}{g_{\theta\theta}}}{\biggr )}
 =2\pi{\biggl (}1- \frac{\kappa}{2} {\biggl |}\frac{d {\cal G}}{dx}
 {\biggl |}_{x_\mathrm{n}}{\biggr )}\:.
 \label{def-N}
 \end{equation}
 Repeating the procedure, this time for the south pole,
 $x_\mathrm{s}$, we conclude that the conical singularity at
 this pole can also be avoid if
\begin{equation}
 \delta_\mathrm{s} \equiv 2\pi{\biggl (}1- \frac{\kappa}{2}
 {\biggl |}\frac{d{\cal G}}{dx}
 {\biggl |}_{x_\mathrm{s}}{\biggr )}=0\:.
 \label{def-S}
 \end{equation}
The so far arbitrary parameter $\kappa$ introduced in Eq.
(\ref{ang}) plays its important role here. Indeed, if we choose
\begin{equation}
\kappa^{-1}=\frac{1}{2}{\biggl |}\frac{d{\cal G}}{dx}
 {\biggl |}_{x=x_\mathrm{s}}\:,
\label{k-s}
 \end{equation}
 Eq. (\ref{def-S}) is satisfied. However, since we only have
a single constant $\kappa$ at our disposal and this has been fixed
to remove the conical singularity at the south pole, we conclude
that the conical singularity will be present at the north pole.
There is another alternative.  We can choose instead
$2\kappa^{-1}=|d_x {\cal G}|_{x=x_\mathrm{n}}$ (where $d_x$ means
derivative in order to $x$) and by doing so we avoid the deficit
angle at the north pole and leave a conical singularity at the
south pole. In section \ref{sec:Phys_Interp} we will see that in
the extended Kruskal solution the north pole points towards the
other black hole, while the south pole points towards  infinity.
The first choice of $\kappa$ corresponds to a strut between the
black holes while the alternative choice corresponds to two
strings from infinity into each black hole. We leave the
discussion on the physical nature of the conical singularities and
on the two possible choices for the value of $\kappa$ to section
\ref{sec:PI-strut}. There is a small number of very special cases
for which the very particular condition,
 $|d_x{\cal G}|_{x_\mathrm{n}}= |d_x{\cal G}|_{x_\mathrm{s}}$
is verified. In these special cases, the solution is free of
conical singularities. They will be mentioned bellow.

Since we have managed to put ${\cal G}(x)$ in a form equal to
\cite{KW}, we can now, following \cite{KW} closely, describe the
behavior of ${\cal G}(x)$ for different values of the parameters
$A$, $m$, and $q$. We can divide this discussion in three cases.

{\it 1. Massless uncharged solution} ($m =0$, $q=0$): in this
case, we have $x=\cos \theta$, ${\cal G}=1-x^2=\sin^2 \theta$, and
$\kappa=1$. The angular surface $\Sigma$ is a sphere and this is a
particular case for which both the north and south poles are free
of conical singularities.

{\it 2. Massive uncharged solution} ($m>0$, $q=0$): the massive
uncharged case  must be divided into $mA<3^{-3/2}$, and $mA \geq
3^{-3/2}$. When $mA<3^{-3/2}$, ${\cal G}(x)$ has three roots and,
as justified above, we require $x$ to lie between the two roots
for which ${\cal G}(x)\geq 0$. In doing so we maintain the metric
with the correct signature and have an angular surface $\Sigma$
which is compact. Setting the value of $\kappa$ given in Eq.
(\ref{k-s}) one avoids the conical singularity at the south pole
but leave one at the north pole. When $mA \geq 3^{-3/2}$, $\Sigma$
is an open angular surface. For this reason, onwards we will
analyze only the case $mA<3^{-3/2}$.

{\it 3. Massive charged solution} ($m>0$, $q\neq0$): for a general
massive charged solution, depending on the values of the
parameters $A$, $m$ and $q$, ${\cal G}(x)$ can be positive in a
single compact interval, $]x_\mathrm{s},x_\mathrm{n}[$, or in two
distinct compact intervals, $]x'_\mathrm{s},x'_\mathrm{n}[$ and
$]x_\mathrm{s},x_\mathrm{n}[$, say. In this latter  case we will
work only with the interval $[x_\mathrm{s},x_\mathrm{n}]$ (say)
for which the charged solutions reduce to the uncharged solutions
when $q=0$. These solutions have a conical singularity at one of
the poles. The only massive charged solutions that are totally
free of conical singularities are those which satisfy the
particular conditions $m=|q|$ and $mA>1/4$. This indicates that in
this case the AdS C-metric is an AdS black hole written in an
accelerated coordinate frame. In the massless charged solution
($m= 0$ and $q\neq 0$), ${\cal G}(x)$ is an even function, has two
symmetric roots and is positive between them. The angular surface
$\Sigma$ is therefore compact and there are no conical
singularities at both poles. Once again, this suggests that the
solution is written in an accelerated coordinate frame.

\subsection{Coordinate ranges}

In this section we analyze  the important issue of the coordinate
ranges. Rewritten in terms of the new coordinates introduced in
Eq. (\ref{r}) and Eq. (\ref{ang}), the AdS C-metric is given by
\begin{equation}
 d s^2 = r^2 [-{\cal F}(y)dt^2+
 {\cal F}^{-1}(y)dy^2+d\theta^2 + \kappa^2{\cal{G}}(x_{(\theta)})d\phi^2]\:,
 \label{AdS C-metric}
 \end{equation}
where ${\cal F}(y)$ and ${\cal{G}}(x_{(\theta)})$ are given by Eq.
(\ref{FG}). The time coordinate $t$ can take any value from the
interval $]-\infty,+\infty[$ and $\phi$ ranges between $[0,2\pi]$.
As we saw in section \ref{sec:CurvSing}, when $m$ or $q$ are not
zero there is a curvature singularity at $r=0$. Therefore, we
restrict the radial coordinate to the range $[0,+\infty[$. On the
other hand, in section \ref{sec:ConSing} we have decided to
consider only the values of $A$, $m$, and $q$ for which ${\cal
G}(x)$ has at least two real roots, $x_\mathrm{s}$ and
$x_\mathrm{n}$ (say) and have demanded $x$ to belong to the range
$[x_\mathrm{s},x_\mathrm{n}]$ where ${\cal G}(x)\geq 0$. By doing
this we guarantee that the metric has the correct signature
$(-+++)$ and that the angular surfaces $\Sigma$ ($t=$constant and
$r=$constant) are compact. From $Ar=(x+y)^{-1}$ we then conclude
that $y$ must belong to the range $-x\leq y < +\infty$. Indeed,
$y=-x$ corresponds to $r=+\infty$, and $y=+\infty$ to $r=0$. Note
however, that when both $m$ and $q$ vanish there are no
restrictions on the ranges of $r$ and $y$ (i.e., $-\infty < r <
+\infty$ and $-\infty < y < +\infty$) since in this case there is
no curvature singularity at the origin of $r$ to justify the
constraint on the coordinates.

\section{\label{sec:PD} Causal Structure of the A\lowercase{d}S
C-metric}

In this section we analyze the causal structure of the solution.
As occurs with the original flat C-metric \cite{KW,AshtDray}, the
original AdS C-metric, Eq. (\ref{AdS C-metric}), is not
geodesically complete. To obtain the maximal analytic spacetime,
i.e., to draw the Carter-Penrose diagrams we will introduce the
usual null Kruskal coordinates.

We now look carefully to the AdS C-metric, Eq. (\ref{AdS
C-metric}), with ${\cal F}(y)$ given by  Eq. (\ref{FG}). We first
notice that, contrarily to what happens in the $\Lambda\geq 0$
background where the causal structure and physical nature of the
corresponding C-metric is independent of the relation between the
acceleration $A$ and $\ell \equiv \sqrt{3/|\Lambda}|$, in the
$\Lambda < 0$ case we must distinguish and analyze separately the
cases $A>1/\ell$, $A=1/\ell$ and $A<1/\ell$. Later, in section
\ref{sec:Phys_Interp}, we will justify physically the reason for
this distinction. The mathematical reason for this difference is
clearly identified by setting $m=0$ and $q=0$ in Eq. (\ref{FG})
giving ${\cal F}(y)=y^2-[1-1/(\ell^2A^2)]$. Since the horizons of
the solution are basically given by the real roots of ${\cal
F}(y)$, we
 conclude that we have to treat separately the cases
(A) $A>1/\ell$ for which ${\cal F}(y)$ can have two real roots,
(B) $A=1/\ell$ for which $y=0$ is double root and (C) $A<1/\ell$
for which ${\cal F}(y)$ has no real roots (see discussion in the
text of Fig. \ref{g1}). We will consider each of these three cases
separately in  subsections \ref{sec:PD A} and \ref{sec:PI.2-BH}
($A>1/\ell$ case), \ref{sec:PD B} and \ref{sec:PI.1-BH}
($A=1/\ell$ case), and \ref{sec:PD C} and \ref{sec:PI.1-BH.1}
($A<1/\ell$ case). The description of the solution depends
crucially on the values of $m$ and $q$. In each subsection, we
will consider the three most relevant solutions, namely: {\it 1.
Massless uncharged solution} ($m =0$, $q=0$), {\it 2. Massive
uncharged solution} ($m>0$, $q=0$), and {\it 3. Massive charged
solution} ($m>0$, $q\neq0$).

\subsection{\label{sec:PD A}
 Causal Structure of the $\bm{A>1/\ell}$ solutions}

\subsubsection{\label{sec:PD A.1}
 \textbf{Massless uncharged solution ($\bm{m=0, q=0}$)}}

\begin{figure}[bh]
\includegraphics*[height=2.2in]{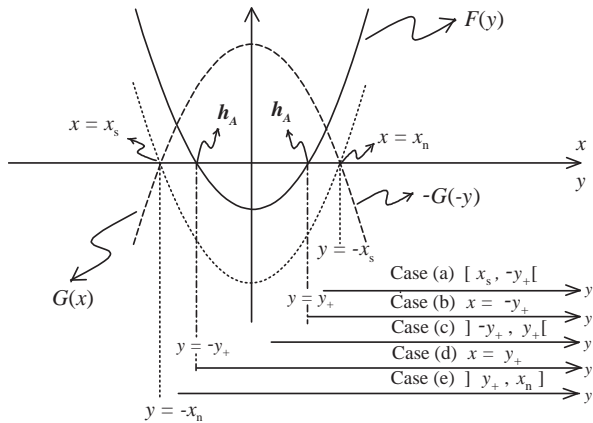}
\caption{\label{g1}
 Shape of ${\cal G}(x)$ and ${\cal F}(y)$ for the
 $A>1/\ell\,,\,m=0\,$ and $q=0$ C-metric studied in sections
\ref{sec:PD A.1} and \ref{sec:PI.2-BH}. The allowed range of $x$
is between $x_\mathrm{s}=-1$ and $x_\mathrm{n}=+1$ where ${\cal
G}(x)$ is positive and compact. The permitted range of $y$ depends
on the angular direction $x$ ($-x\leq y < +\infty$) and is
sketched for the five cases (a)-(e) discussed in the text. The
presence of an accelerated horizon is indicated by $h_A$. [For
completeness we comment here on two other cases not represented in
the figure but analyzed in the text: for $A=1/\ell\,,\,m=0\,$ and
$q=0$ (this case is studied in sections \ref{sec:PD B.1} and
\ref{sec:PI.1-BH}), ${\cal F}(y)$ is zero at its minimum and
positive elsewhere.  For $A<1/\ell\,,\,m=0\,$ and $q=0$ (this case
is studied in sections \ref{sec:PD C.1} and \ref{sec:PI.1-BH.1}),
${\cal F}(y)$ is always positive and only case (a) survives.]
 }
\end{figure}

In this case we have
\begin{eqnarray}
 {\cal F}(y) = y^2-y_+^2 \;\;\;\;\;\;\mathrm{with}\;\;\;\;\;\;
 y_+=\sqrt{1-\frac{1}{\ell^2A^2}} \:,
 \label{F1}
 \end{eqnarray}
and $x \in [x_\mathrm{s}=-1,x_\mathrm{n}=+1]$, $x=\cos \theta$,
${\cal G}=1-x^2=\sin^2 \theta$ and $\kappa=1$. The shapes of
${\cal F}(y)$ and ${\cal G}(x)$ are represented in Fig. \ref{g1}.

 The angular surfaces $\Sigma$ ($t=$constant and
$r=$constant) are spheres and both the north and south poles are
free of conical singularities. The origin of the radial
coordinate, $r=0$, has no curvature singularity and therefore both
$r$ and $y$ are in the range $]-\infty,+\infty[$. However, in the
general case, where $m$ or $q$ are non-zero, there is a curvature
singularity at $r=0$. Since the discussion of the present section
is only a preliminary to that of the massive general case,
following \cite{AshtDray}, the origin $r=0$ will treat as if it
had a curvature singularity and thus we admit that $r$ belongs to
the range $[0,+\infty[$ and $y$ lies in the region $-x\leq y <
+\infty$. We leave a discussion on the extension to negative
values of $r$ to section \ref{sec:PI.2-BH}.

The general procedure to draw the Carter-Penrose diagrams is as
follows. First, we make use of the null condition
$g_{\mu\nu}k^{\mu}k^{\nu}=0$ (where $k^{\mu}$ is a geodesic
tangent) to introduce the advanced and retarded
Finkelstein-Eddington null coordinates,
\begin{eqnarray}
 u=t-y_* \:;    \;\;\;\;\;\;   v=t+y_* \:,
 \label{uv}
 \end{eqnarray}
where the tortoise coordinate is
\begin{eqnarray}
y_*=\int {\cal F}^{-1}dy=\frac{1}{2y_+} \ln{{\biggl
|}\frac{y-y_+}{y+y_+}{\biggl |}}\:.
 \label{y*}
 \end{eqnarray}
and both $u$ and $v$ belong to the range $]-\infty,+\infty[$. In
these coordinates the metric is given by
\begin{equation}
 d s^2 = r^2 [-{\cal F}dudv+
 d\theta^2 + \sin^2\!\theta \, d\phi^2]\:.
 \label{A.1.1}
 \end{equation}
 The metric still has coordinate
 singularities at the roots of ${\cal F}$. To overcome this
 unwanted feature we have to introduce Kruskal
 coordinates. Now, due to the lower restriction on the value of $y$
 ($-x\leq y$), the choice of the Kruskal coordinates (and therefore the
 Carter-Penrose diagrams) depends on the angular direction $x$ we
 are looking at. In fact, depending on the value of $x$, the
 region accessible to $y$ might contain two, one or zero roots of
 ${\cal F}$ (see Fig. \ref{g1}) and so the solution may have two,
 one or zero horizons, respectively. This angular dependence of the
 causal diagram is not new. The Schwarzschild and
  Reissner-Nordstr\"{o}m solutions being spherically symmetric do
 not present this feature but, in the Kerr solution, the
 Carter-Penrose diagram along the pole direction is different
 from the diagram along the equatorial direction. Such a dependence
 occurs also in the flat C-metric \cite{KW}. Back again to the
 AdS C-metric, we have to consider separately five distinct sets of
 angular directions, namely (a) $x_\mathrm{s}\leq x <-y_+$,  (b) $x =-y_+$,  (c)
 $ -y_+ < x <y_+$,  (d) $x=+y_+$ and (e)
 $y_+ < x \leq x_\mathrm{n}$, where $x_\mathrm{s}=-1$ and
 $x_\mathrm{n}=+1$ (see Fig. \ref{g1}).

\vspace{0.3 cm} (a) $x_\mathrm{s}\leq x <-y_+$: within this
interval of the angular direction, the restriction on the range of
$y$, $-x\leq y < +\infty$, implies that the function ${\cal F}(y)$
is always positive in the accessible region of $y$ (see Fig.
\ref{g1}), and thus the solution has no horizons. Introducing the
null coordinates defined in Eq. (\ref{uv}) followed by the Kruskal
coordinates $u'=-e^{-y_+ u}<0$ and $v'=+e^{+y_+ v}>0$ gives
$u'v'=-e^{2y_+y_*}=-(y-y_+)/(y+y_+)<0$, and  Eq. (\ref{A.1.1})
becomes
\begin{eqnarray}
 d s^2 =  r^2 {\biggl [}-\frac{(y+y_+)^2}{y_+^2}du'dv'+
 d\theta^2 + \sin^2\!\theta \,d\phi^2 {\biggr ]} \:,
 \label{A.1.2}
 \end{eqnarray}
where $y$ and $r=A^{-1}(x+y)^{-1}$ are regarded as functions of
$u'$ and $v'$,
\begin{eqnarray}
 y=y_+\frac{1-u'v'}{1+u'v'}\:, \;\;\;\;
 r=\frac{1}{A}\frac{1+u'v'}{(y_+ +x)-u'v'(y_+ -x)} \:.
 \label{y,r}
 \end{eqnarray}
Now, let us find the values of the product $u'v'$ at $r=0$ and
$r=+\infty$,
\begin{eqnarray}
 \lim_{r \to 0} u'v'=-1\:, \;\;\;\;\;
 \lim_{r \to +\infty} u'v'=\frac{y_+ + x}{y_+ - x}<0 \;
 \mathrm{and \; finite} \:. \nonumber \\
 \label{lim u'v'}
 \end{eqnarray}
So, for $x_\mathrm{s}\leq x <-y_+$, the original massless
uncharged AdS C-metric is described by  Eq. (\ref{A.1.2})
subjected to the following coordinates ranges,
\begin{eqnarray}
 \hspace{-0.5cm} & & \hspace{-0.5cm}
 0 \leq \phi < 2\pi\:, \;\;\; -1 \leq x \leq +1 \:,\;\;\; u'<0\:,
 \;\;\; v'>0 \:,\;\;\;          \\
 \hspace{-0.5cm} & & \hspace{-0.5cm}
  -1\leq u'v'<\frac{y_+ + x}{y_+ - x}  \:.
 \label{ranges u'v'}
 \end{eqnarray}
 This spacetime is however geodesically incomplete. To obtain the
 maximal analytical extension one allows the Kruskal coordinates
 to take also the values $u'\geq 0$ and $v'\leq 0$ as long as
  Eq. (\ref{ranges u'v'}) is satisfied.

\begin{figure}
\includegraphics*[height=4.0in]{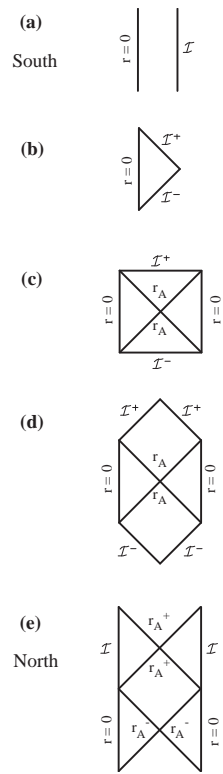}
   \caption{\label{Fig-a1}
Carter-Penrose diagrams of cases (a)-(e) discussed in the text of
section \ref{sec:PD A.1} concerning the $A>1/\ell\,,\,m=0\,,\,q=0$
C-metric. Case (a) describes the solution seen from the vicinity
of the south pole, case (c) applies to the equatorial vicinity,
and case (e) describes the solution seen from the vicinity of the
north pole. An accelerated horizon is represented by $r_A$, and
${\cal I}^-$ and ${\cal I}^+$ represent respectively the past and
future infinity ($r=+\infty$). $r=0$ corresponds to $y=+\infty$
and $r=+\infty$ corresponds to $y=-x$.
 }
\end{figure}

Finally, to construct the Carter-Penrose diagram one has to define
the Carter-Penrose coordinates by the usual arc-tangent functions
of $u'$ and $v'$: ${\cal{U}}=\arctan u'$ and ${\cal{V}}=\arctan
v'$, that bring the points at infinity into a finite position. In
general, to find what kind of curve describes the lines $r=0$ or
$r=+\infty$ one has to take the limit of $u'v'$ as $r\rightarrow
0$ (in the case of $r=0$) and the limit of $u'v'$ as $r\rightarrow
+\infty$ (in the case of $r=+\infty$). If this limit is $0$ or
$\infty$ the corresponding line is mapped into a curved null line.
If the limit is $-1$, or a negative and finite constant, the
corresponding line is mapped into a curved timelike line and
finally, when the limit is $+1$, or a positive and finite
constant, the line is mapped into a curved spacelike line. The
asymptotic lines are drawn as straight lines although in the
coordinates ${\cal{U}}$ and ${\cal{V}}$ they should be curved
outwards, bulged. It is always possible to change coordinates so
that the asymptotic lines are indeed straight lines. So, from Eq.
(\ref{lim u'v'}) we draw the Carter-Penrose diagram sketched in
Fig. \ref{Fig-a1}.(a). There are no horizons and both $r=0$ and
$r=+\infty$ (${\cal I}$) are timelike lines.

\vspace{0.3 cm}
 (b) $x =-y_+$: for this particular angular
direction, $y$ is restricted to be on $+y_+ \leq y < +\infty$ and
${\cal F}(y)$ is always positive except at $y=+y_+$ (which
corresponds to $r=+\infty$) where it is zero (see Fig. \ref{g1}).
Therefore, the solution has no horizon and the Kruskal
construction is similar to the one described above in case (a).
The only difference is that now $\lim_{r \to +\infty} u'v'=0$ and
thus $r=+\infty$ (${\cal I}$) is represented by a null line in the
Carter-Penrose diagram which is shown in Fig. \ref{Fig-a1}.(b).

\vspace{0.3 cm}
 (c) $-y_+ < x <y_+$: the demand that $y$ must belong to the range
$[-x;+\infty[$ implies, for this range of the angular direction,
that we have a region I, $-x\leq y < +y_+$, where ${\cal F}(y)$ is
negative and a region II, $+y_+<y<+\infty$, where ${\cal F}(y)$ is
positive (see Fig. \ref{g1}). There is a single Rindler-like
acceleration horizon ($r_A$) at $y=+y_+$, so called because it is
is absent when $A=0$ and present even when $m=0$ and $q=0$. In
region I one sets the Kruskal coordinates $u'=+e^{-\alpha u}$ and
$v'=+e^{+\alpha v}$ so that
 $u'v'=+e^{2\alpha y_*}$.
In region II one defines $u'=-e^{-\alpha u}$ and $v'=+e^{+\alpha
v}$ in order that $u'v'=-e^{2\alpha y_*}$. We set $\alpha\equiv
y_+$. Thus, in both regions the product $u'v'$ is given by
\begin{eqnarray}
u'v'=-\frac{y-y_+}{y+y_+} \:,
 \label{u'v'}
\end{eqnarray}
and Eq. (\ref{A.1.1}) expressed in terms of the Kruskal
coordinates is given by
\begin{eqnarray}
 d s^2 &=& r^2 {\biggl [}\frac{1}{y_+^2}\frac{{\cal F}}{u'v'}du'dv'+
 d\theta^2 + \sin^2\!\theta \,d\phi^2 {\biggr ]}    \label{F/u'v'} \\
      &=&  r^2 {\biggl [}-\frac{(y+y_+)^2}{y_+^2}du'dv'+
 d\theta^2 + \sin^2\!\theta \,d\phi^2 {\biggr ]}. \nonumber \\
 \label{A.1.3}
 \end{eqnarray}
The Kruskal coordinates in both regions were chosen in order to
obtain a negative value for the factor ${\cal F}/(u'v')$, which
appears in the metric coefficient $g_{u'v'}$. The value of
constant $\alpha$ was selected in order that the limit of ${\cal
F}/(u'v')$ as $y \to y_+$ stays finite and different from zero. By
doing this, we have removed the coordinate singularity that was
present at the root $y_+$ of ${\cal F}$ [see Eq. (\ref{A.1.1})].
So, the metric is now well-behaved in the whole range $-x\leq y
<+\infty$ or $0\leq r<+\infty$. The coordinates $y$ and $r$ are
expressed as functions of $u'$ and $v'$ by Eq. (\ref{y,r}) and at
the edges of the interval allowed to $r$, the product $u'v'$ takes
the values
\begin{eqnarray}
 \lim_{r \to 0} u'v'=-1\:, \;\;\;\;\;
 \lim_{r \to +\infty} u'v'=\frac{y_+ + x}{y_+ - x}>0 \;
 \mathrm{and \; finite} \:. \nonumber \\
 \label{lim u'v'.2}
 \end{eqnarray}
Once again, the maximal analytical extension is achieved by
allowing the Kruskal coordinates $u'$ and $v'$ to take all the
values on the range $]-\infty;+\infty[$, as soon as the condition
 $-1\leq u'v'<(y_+ + x)/(y_+ - x)$ is satisfied.
 The Carter-Penrose diagram for this range of the angular
 direction is drawn in Fig. \ref{Fig-a1}.(c). $r=0$ is represented by a timelike
line while $r=+\infty$ (${\cal I}$) is a spacelike line. The two
mutual perpendicular straight null lines at $45^{\rm o}$,
$u'v'=0$, represent the accelerated horizon at $y_A=+y_+$ or
$r_A=[A(x+y_+)]^{-1}$.

\vspace{0.3 cm}

 (d) $x = +y_+$: in this particular direction, the region accessible to
$y$ is $-y_+\leq y < +\infty$. ${\cal F}(y)$ is negative in region
I, $-y_+ < y < y_+$ and positive in region II, $y > y_+$. It is
zero at $y=+y_+$ where is located the only horizon ($r_A$) of the
solution and ${\cal F}(y)$ vanishes again at $y=-y_+$ which
corresponds to $r=+\infty$ (see Fig. \ref{g1}). The Kruskal
construction follows directly the procedure described in case (c).
The only difference is that now $\lim_{r \to +\infty}
u'v'=+\infty$ and thus the $r=+\infty$ line (${\cal I}$) is now
represented by a null line in the Carter-Penrose diagram which is
shown in Fig. \ref{Fig-a1}.(d).

\vspace{0.3 cm}
 (e) $y_+ < x \leq x_\mathrm{n}$: the region accessible to $y$ must
be separated into three regions (see Fig. \ref{g1}). In region I,
$-x < y < -y_+$, ${\cal F}(y)$ is positive; in region II, $-y_+ <
y < +y_+$, ${\cal F}(y)$ is negative and finally in region III,
$y>+y_+$, ${\cal F}(y)$ is positive again. We have two
Rindler-like acceleration horizons, more specifically, an outer
horizon at $y=-y_+$ or $r_A^+ =[A(x-y_+)]^{-1}$ and an inner
horizon at $y=+y_+$ or $r_A^-=[A(x+y_+)]^{-1}$. Therefore one must
introduce a Kruskal coordinate patch around each of the horizons.
The first patch constructed around $-y_+$ is valid for $-x \leq y
< +y_+$ (thus, it includes regions I and II). In region I we
define $u'=-e^{+\alpha_- u}$ and $v'=+e^{-\alpha_- v}$ so that
$u'v'=-e^{-2\alpha_- y_*}$. In region II one defines
$u'=+e^{\alpha_- u}$ and $v'=+e^{-\alpha_- v}$ in order that
$u'v'=+e^{-2\alpha_- y_*}$. We set $\alpha_-\equiv y_+$. Thus, in
both regions, I and II, the product $u'v'$ is given by
\begin{eqnarray}
u'v'=-\frac{y+y_+}{y-y_+} \:,
 \label{u'v'.2}
\end{eqnarray}
and  Eq. (\ref{A.1.1}) expressed in terms of the Kruskal
coordinates is given by
\begin{eqnarray}
 d s^2 = r^2 {\biggl [}-\frac{(y-y_+)^2}{y_+^2}du'dv'+
 d\theta^2 + \sin^2\!\theta \,d\phi^2 {\biggr ]} \:,
 \label{A.1.4}
 \end{eqnarray}
which is regular in this patch $-x \leq y < +y_+$ and, in
particular, it is regular at the root $y=-y_+$ of ${\cal F}(y)$.
However, it is singular at the second root, $y=+y_+$, of
 ${\cal F}(y)$. To regularize the metric around $y=+y_+$, one has
 to introduce new Kruskal coordinates for the second patch which
is built around $y_+$ and is valid for $-y_+ < y < +\infty$ (thus,
it includes regions II and III). In region II we set
$u'=+e^{-\alpha_+ u}$ and $v'=+e^{+\alpha_+ v}$ so that
$u'v'=+e^{+2\alpha_+ y_*}$. In region III one defines
$u'=-e^{-\alpha_+ u}$ and $v'=+e^{+\alpha_+ v}$ in order that
$u'v'=-e^{+2\alpha_+ y_*}$. We set $\alpha_+\equiv y_+$. Thus, in
both regions, II and III, the product $u'v'$ is given by
\begin{eqnarray}
u'v'=-\frac{y-y_+}{y+y_+} \:,
 \label{u'v'.3}
\end{eqnarray}
and, in this second Kruskal patch,  Eq. (\ref{A.1.1}) is given by
\begin{eqnarray}
 d s^2 = r^2 {\biggl [}-\frac{(y+y_+)^2}{y_+^2}du'dv'+
 d\theta^2 + \sin^2\!\theta \,d\phi^2 {\biggr ]} \:,
 \label{A.1.5}
 \end{eqnarray}
which is regular in $ y > -y_+$ and, in particular, at the second
root $y=+y_+$ of ${\cal F}(y)$. Once again, in both patches, the
Kruskal coordinates were chosen in order to obtain a factor ${\cal
F}/(u'v')$ negative [see Eq. (\ref{F/u'v'})]. The values of
constants $\alpha_-$ and $\alpha_+$ were selected in order that
the limit of ${\cal F}/(u'v')$ as $y \to \mp y_+$ stays finite and
different from zero. To end the construction of the Kruskal
diagram of this solution with two horizons, the two patches have
to be joined together in an appropriate way first defined by
Carter in the Reissner-Nordstr\"{o}m solution.

 From Eq. (\ref{u'v'.3}) and Eq. (\ref{u'v'.2}) we find the values of
product $u'v'$ at the edges $r=0$ and $r=+\infty$ of the radial
coordinate,
\begin{eqnarray}
 \lim_{r \to 0} u'v'=-1\:, \;\;\;\;\;
 \lim_{r \to +\infty} u'v'=\frac{y_+ - x}{y_+ + x}<0 \;
 \mathrm{and \; finite} \:. \nonumber \\
 \label{lim u'v'.4}
 \end{eqnarray}
and conclude that both $r=0$ and $r=+\infty$ (${\cal I}$) are
represented by timelike lines in the Carter-Penrose diagram
sketched in Fig. \ref{Fig-a1}.(e). The two accelerated horizons of
the solution are both represented as perpendicular straight null
lines at $45^{\rm o}$ ($u'v'=0$).

\subsubsection{\label{sec:PD A.2}
 \textbf{Massive uncharged solution ($\bm{m >0}$, $\bm{q=0}$)}}

\begin{figure} [b]
\includegraphics*[height=2.2in]{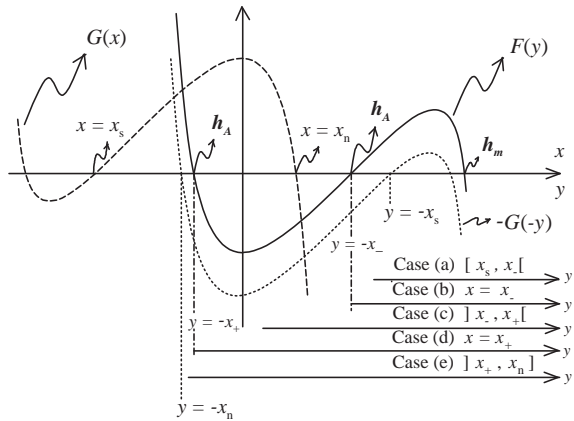}
\caption{\label{g2}
 Shape of ${\cal G}(x)$ and ${\cal F}(y)$ for the
 $A>1/\ell, mA<3^{-3/2}$, and $q=0$ C-metric studied in sections
\ref{sec:PD A.2} and \ref{sec:PI.2-BH}. The allowed range of $x$
is between $x_\mathrm{s}$ and $x_\mathrm{n}$ where ${\cal G}(x)$
is positive and compact. The permitted range of $y$ depends on the
angular direction $x$ ($-x\leq y < +\infty$) and is sketched for
the five cases (a)-(e) discussed in the text. The presence of an
accelerated horizon is indicated by $h_A$ and the
Schwarzschild-like horizon by $h_m$. [For completeness we comment
on two other cases not represented in the figure: for $A=1/\ell,
mA<3^{-3/2}$ and $q=0$ (this case is studied in sections
\ref{sec:PD B.2} and \ref{sec:PI.1-BH}), ${\cal F}(y)$ is zero at
its local minimum. For $A<1/\ell, mA<3^{-3/2}$ and $q=0$ (this
case is studied in sections \ref{sec:PD C.2} and
\ref{sec:PI.1-BH.1}), the local minimum of ${\cal F}(y)$ is
positive and only case (a) survives. For $mA=3^{-3/2}$, ${\cal
G}(x)$ is zero at its local minimum on the left and for
$mA>3^{-3/2}$ ${\cal G}(x)$ is positive between $-\infty$ and
$x_\mathrm{n}$. These two last cases are not studied in the text.]
 }
\end{figure}

Now that the causal structure of the AdS C-metric with $m=0$ and
$q=0$ has been studied, the construction of the Carter-Penrose
diagrams for the $m >0$ case follows up directly. As has justified
in detail in section \ref{sec:ConSing}, we will consider only the
case with small mass or acceleration, i.e., we require
$mA<3^{-3/2}$, in order to have  compact angular surfaces (see
discussion on the text of Fig. \ref{g2}). We also demand $x$ to
belong to the range $[x_\mathrm{s},x_\mathrm{n}]$ (see Fig.
\ref{g2}) where ${\cal G}(x)\geq 0$ and such that $x_\mathrm{s}
\to -1$ and $x_\mathrm{n} \to +1$ when $mA \to 0$. By satisfying
the two above conditions we endow the $t=$constant and
$r=$constant surfaces with the topology of a compact surface.

\begin{figure}[hb]
\includegraphics*[height=13cm]{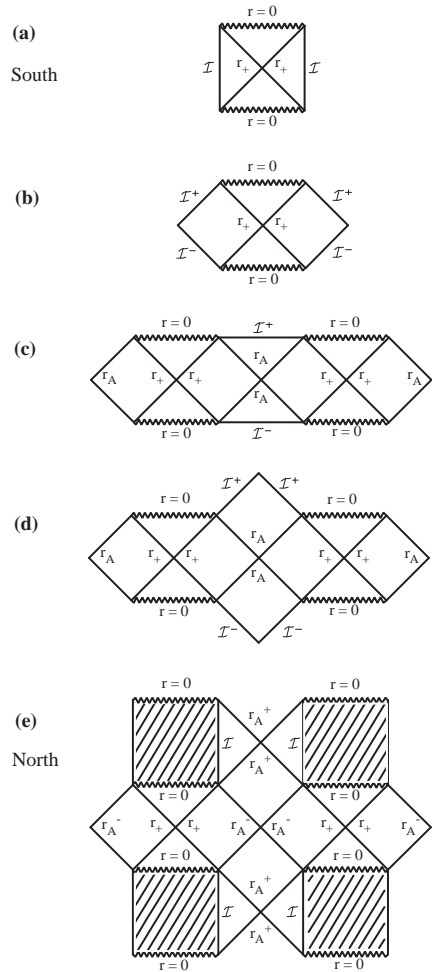}
   \caption{\label{Fig-a2}
Carter-Penrose diagrams of cases (a)-(e) discussed in the text of
section \ref{sec:PD A.2} concerning the $A>1/\ell, mA<3^{-3/2}$,
and $q=0$ C-metric. Case (a) describes the solution seen from the
vicinity of the south pole, case (c) applies to the equatorial
vicinity, and case (e) describes the solution seen from the
vicinity of the north pole. The zigzag line represents a curvature
singularity, an accelerated horizon is represented by $r_A$, the
Schwarzschild-like horizon is sketched as $r_+$.  $r=0$
corresponds to $y=+\infty$ and $r=+\infty$ (${\cal I}$)
corresponds to $y=-x$. The hatched region does not belong to the
solution. In diagrams (c)-(e) we have to glue indefinitely copies
of the represented figure in the left and right sides of it. In
diagram (e) a similar gluing must be done in the top and bottom
regions.
 }
\end{figure}

The technical procedure to obtain the Carter-Penrose diagrams is
similar to the one described along section \ref{sec:PD A.1}. In
what concerns the physical conclusions, we will see that the
essential difference is the presence of an extra horizon, a
Schwarzschild-like horizon ($r_+$), due to the non-vanishing mass
parameter, in addition to the accelerated Rindler-like horizon
($r_A$) which has due to non-vanishing $A$. Another important
difference, as stated in section \ref{sec:CurvSing}, is the
presence of a curvature singularity at $r=0$ and the existence of
a conical singularity at the north pole (see section
\ref{sec:ConSing}).

 Once more the Carter-Penrose diagrams
depend on the angular direction we are looking at (see Fig.
\ref{g2}). We have to analyze separately five distinct cases,
namely (a)
 $x_\mathrm{s}\leq x < x_-$, (b) $x =x_-$,  (c)
 $ x_- < x < x_+$, (d) $x=x_+$ and (e)
 $x_+ < x \leq x_\mathrm{n}$, which are the massive counterparts of
 cases (a)-(e) that were considered in section
  \ref{sec:PD A.1}.
 When $m \to 0$ we have $x_\mathrm{s} \to -1$, $x_\mathrm{n} \to +1$,
$x_- \to -y_+$ and $x_+ \to +y_+$.

\vspace{0.3 cm}
 (a) $x_\mathrm{s}\leq x <x_-$:
the Carter-Penrose diagram [Fig. \ref{Fig-a2}.(a)] for this range
of the angular direction has a spacelike curvature singularity at
$r=0$, a timelike line that represents $r=+\infty$ (${\cal I}$)
and a Schwarzschild-like horizon ($r_+$) that was not present in
the $m=0$ corresponding diagram Fig. \ref{Fig-a1}.(a). The diagram
is similar to the one of the AdS-Schwarzschild solution, although
now the curvature singularity has an acceleration $A$, as will be
seen in section \ref{sec:Phys_Interp}.

\vspace{0.3 cm}
 (b) $x =x_-$:
the curvature singularity $r=0$ is also a spacelike line in the
Carter-Penrose diagram [see Fig \ref{Fig-a2}.(b)] and there is a
Schwarzschild-like horizon ($r_+$). The infinity, $r=+\infty$
(${\cal I}$), is represented by a null line. The origin is being
accelerated (see section \ref{sec:Phys_Interp}).

\vspace{0.3 cm}
  (c) $x_- < x <x_+$:
the Carter-Penrose diagram [Fig. \ref{Fig-a2}.(c)] has a more
complex structure that can be divided into left, middle and right
regions. The middle region contains the spacelike infinity (${\cal
I}$) and an accelerated Rindler-like horizon,
$r_A=[A(x-x_-)]^{-1}$, that is already present in the $m=0$
corresponding diagram [see Fig. \ref{Fig-a1}.(c)]. The left and
right regions both contain a spacelike curvature singularity and a
Schwarzschild-like horizon, $r_+$.

\vspace{0.3 cm}
 (d) $x = x_+$:
the Carter-Penrose diagram [Fig. \ref{Fig-a2}.(d)] for this
particular value of the angular direction is similar to the one of
above case (c). The only difference is that $r=+\infty$ (${\cal
I}$) is represented by a null line rather than a spacelike line.

\vspace{0.3 cm}
  (e) $x_+ < x \leq x_\mathrm{n}$:
the Carter-Penrose diagram [Fig. \ref{Fig-a2}.(e)] can again be
divided into left, middle and right regions. The middle region
consists of a timelike line representing $r=+\infty$ (${\cal I}$)
and two accelerated Rindler-like horizons, an inner one
($r_A^-=[A(x-x_-)]^{-1}$) and an outer one
($r_A^+=[A(x-x_+)]^{-1}$), that were already present in the $m=0$
corresponding diagram [Fig. \ref{Fig-a1}.(e)]. The left and right
regions both contain a spacelike curvature singularity and a
Schwarzschild-like horizon ($r_+$).
\subsubsection{\label{sec:PD A.3}
\textbf{Massive charged solution ($\bm{m >0}$, $\bm{q\neq0}$)}}
\begin{figure} [t]
\includegraphics*[height=2.2in]{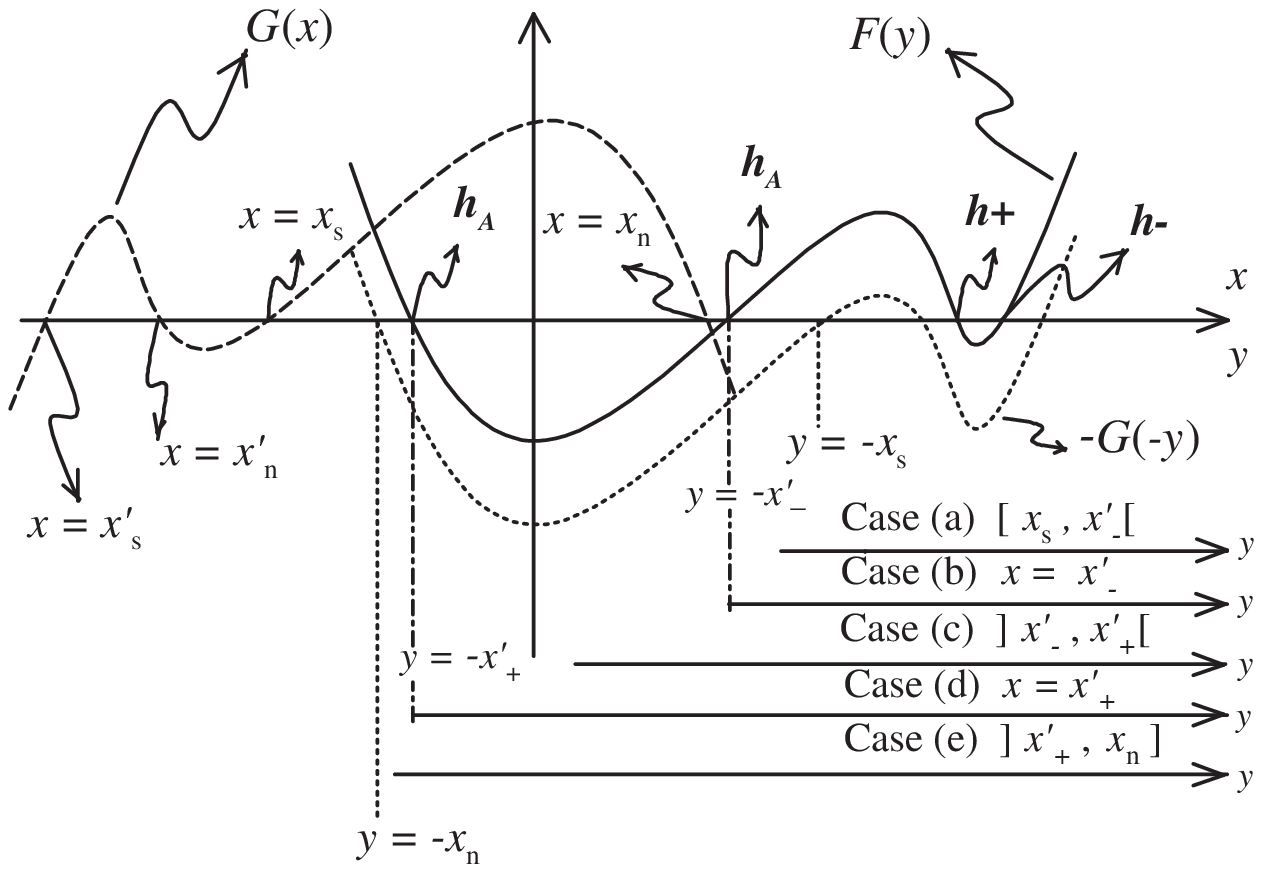}
\caption{\label{g3}
 Shape of ${\cal G}(x)$ and ${\cal F}(y)$ for the non-extremal
charged massive C-metric (with $A>1/\ell$) studied in sections
\ref{sec:PD A.3} and \ref{sec:PI.2-BH}. The allowed range of $x$
is between $x_\mathrm{s}$ and $x_\mathrm{n}$ where ${\cal G}(x)$
is positive and compact. The permitted range of $y$ depends on the
angular direction $x$ ($-x\leq y < +\infty$) and is sketched for
the five cases (a)-(e) discussed in the text. The presence of an
accelerated horizon is indicated by $h_A$ and the inner and outer
charged horizons by $h-$ and $h+$. In the extremal case, $h-$ and
$h+$ superpose each other and in the naked case ${\cal F}(y)>0$ in
the local minimum on the right. [For completeness we comment on
two other cases not represented in the figure: for $A=1/\ell$
(this case is studied in sections \ref{sec:PD B.3} and
\ref{sec:PI.1-BH}), ${\cal F}(y)$ is zero at its local minimum on
the left.  For $A<1/\ell$ (this case is studied in sections
\ref{sec:PD C.3} and \ref{sec:PI.1-BH.1}), the local minimum on
the left of ${\cal F}(y)$ is positive and only case (a) survives.]
 }
\end{figure}

When both the mass and charge parameters are non-zero, depending
on the values of the parameters $A$, $m$ and $q$, ${\cal G}(x)$
can be positive in a single compact interval,
$]x_\mathrm{s},x_\mathrm{n}[$, or in two distinct compact
intervals, $]x'_\mathrm{s},x'_\mathrm{n}[$ and
$]x_\mathrm{s},x_\mathrm{n}[$, say (see Fig. \ref{g3}). In this
latter case we will work only with the interval
$[x_\mathrm{s},x_\mathrm{n}]$ (say) for which the charged
solutions are in the same sector of those we have analyzed in the
last two subsections when $q \to 0$.

Depending also on the values of $A$, $m$ and $q$, the function
${\cal F}(y)$ can have four roots, three roots (one of them
degenerated) or two roots (see the discussion on the text of Fig.
\ref{g3}). As will be seen, the first case describes a pair of
accelerated AdS$-$Reissner-Nordstr\"{o}m (AdS-RN) black holes, the
second case describes a pair of extremal AdS-RN black holes and
the third case describes a pair of naked AdS-RN singularities.

The essential differences between the Carter-Penrose diagrams of
the massive charged solutions and those of the massive uncharged
solutions are: (i) the curvature singularity is now represented by
a timelike line rather than a spacelike line, (ii) excluding the
extremal and naked cases, there are now (in addition to the
accelerated Rindler-like horizon, $r_A$) not one but two extra
horizons, the expected inner ($r_-$) and outer ($r_+$) horizons
associated to the charged character of the solution.

\begin{figure*} [ht]
\includegraphics*[height=19cm]{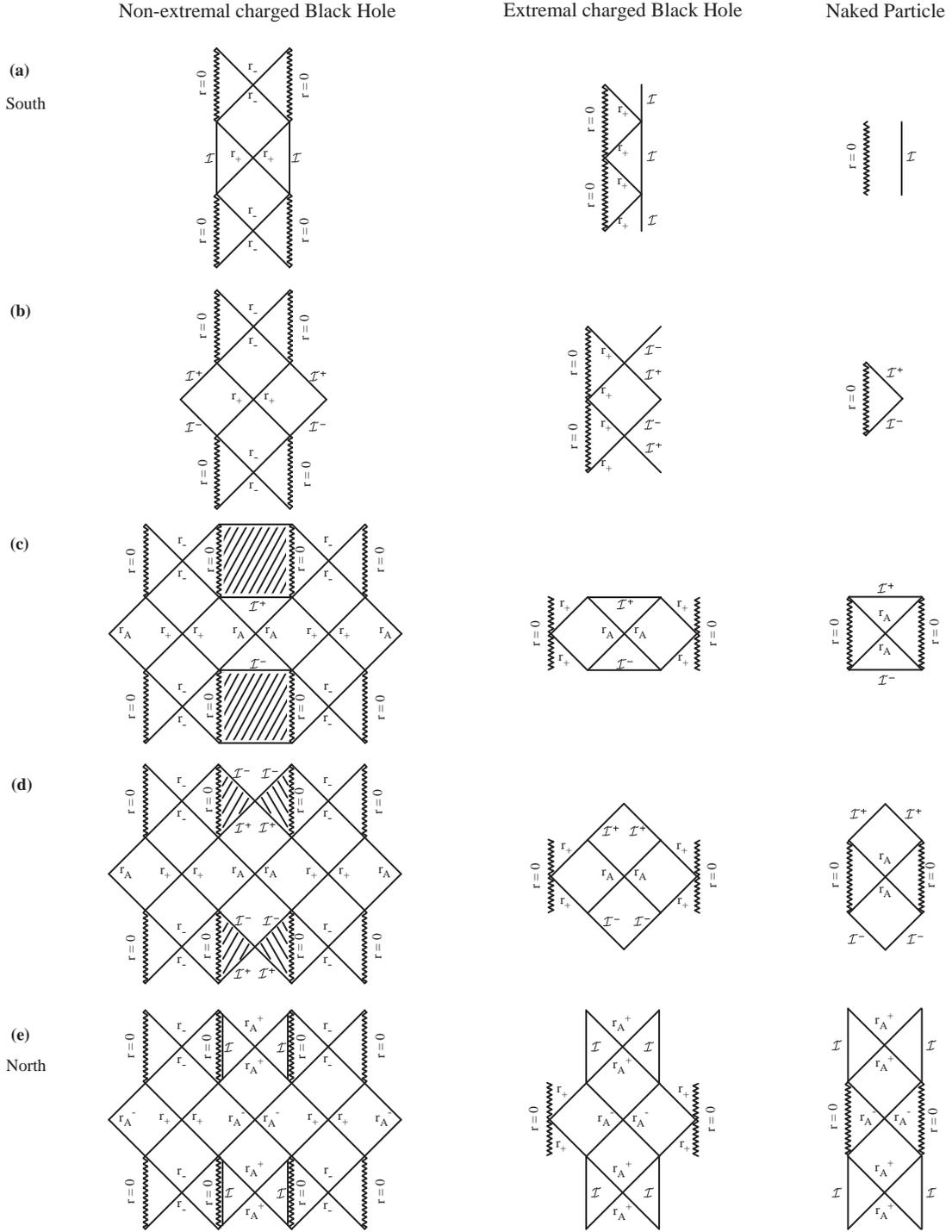}
   \caption{\label{Fig-a3}
Carter-Penrose diagrams of cases (a)-(e) discussed in the text of
section \ref{sec:PD A.3} concerning the charged massive C-metric.
The left column represents the non-extremal case, the middle
column represents the extremal case and the right column
represents the naked charged case. The row (a) describes the
solution seen from the vicinity of the south pole, row (c) applies
to the equatorial vicinity, and row (e) describes the solution
seen from the vicinity of the north pole.
 }
\end{figure*}

 Below, we study the causal structure of the electric or magnetic
 counterparts of cases (a)-(e) discussed in the two
 last sections (see Fig. \ref{g3}), namely (a)
 $x_\mathrm{s}\leq x < x'_-$, (b) $x =x'_-$,  (c)
 $ x'_- < x < x'_+$,  (d) $x=x'_+$ and  (e)
 $x'_+ < x \leq x_\mathrm{n}$.
 When $q \to 0$ we have $x'_- \to x_-$ and $x'_+ \to x_+$.
The Carter-Penrose diagrams are drawn in Fig. \ref{Fig-a3}. In
these diagrams, the left column represents the non-extremal case,
the middle column represents the extremal case and the right
column represents the naked charged case. The row (a) describes
the solution seen from the vicinity of the south pole, row (c)
applies to the equatorial vicinity, and row (e) describes the
solution seen from the vicinity of the north pole. The zigzag line
represents a curvature singularity, an accelerated horizon is
represented by $r_A$, the inner and outer charge associated
horizons are sketched as $r_-$ and $r_+$. ${\cal I}^-$ and ${\cal
I}^+$ represent respectively the past and future infinity
($r=+\infty$). $r=0$ corresponds to $y=+\infty$ and $r=+\infty$
corresponds to $y=-x$. The hatched region does not belong to the
solution. In diagrams (c)-(e) we have to glue indefinitely copies
of the represented figure in the left and right sides of it. In
some of the diagrams, a similar gluing must be done in the top and
bottom regions.

\vspace{0.3 cm}
 (a) $x_\mathrm{s}\leq x <x'_-$:
both the curvature singularity, $r=0$, and $r=+\infty$ (${\cal
I}$) are represented by a spacelike line in the Carter-Penrose
diagram [Fig. \ref{Fig-a3}.(a)]. Besides, in the non-extremal
case, there is an inner horizon ($r_-$) and an outer horizon
($r_+$) associated to the charged character of the solution. In
the extremal case the two horizons $r_-$ and $r_+$ become
degenerate and so there is a single horizon, $r_+$ (say), and in
the naked case there is no horizon. The diagram is similar to the
one of the AdS$-$Reissner-Nordstr\"{o}m solution, although now the
curvature singularity has an acceleration $A$, as will be seen in
section \ref{sec:Phys_Interp}.

\vspace{0.3 cm}
  (b) $x =x'_-$:
the curvature singularity $r=0$ is a spacelike line in the
Carter-Penrose diagram [see Fig. \ref{Fig-a3}.(b)] and $r=+\infty$
(${\cal I}$) is represented by a null line. Again, in the
non-extremal case, there is an inner horizon ($r_-$) and an outer
horizon ($r_+$) associated to the charged character of the
solution. In the extremal case there is a single horizon, $r_+$,
and in the naked case there is no horizon. The origin is being
accelerated (see section \ref{sec:Phys_Interp}).

\vspace{0.3 cm}
 (c) $x'_- < x <x'_+$:
the Carter-Penrose diagram [Fig. \ref{Fig-a3}.(c)] has a complex
structure. As before [see Fig \ref{Fig-a2}.(c)], it can be divided
into left, middle and right regions. The middle region contains
the spacelike infinity (${\cal I}$) and an accelerated
Rindler-like horizon, $r_A=[A(x-x'_-)]^{-1}$, that was already
present in the $m=0\,,\,q=0$ corresponding diagram [see Fig.
\ref{Fig-a1}.(c)]. The left and right regions both contain a
timelike curvature singularity ($r=0$). In addition, these left
and right regions contain, in the non-extremal case, an inner
horizon ($r_-$) and an outer horizon ($r_+$), in the extremal case
they contain is a single horizon ($r_+$), and in the naked case
they have no horizon.

\vspace{0.3 cm}
  (d) $x = x'_+$:
the Carter-Penrose diagram [Fig. \ref{Fig-a3}.(d)] for this
particular value of the angular direction is similar to the one of
above case (c). The only difference is that $r=+\infty$ (${\cal
I}$) is represented by a null line rather than a spacelike line.

\vspace{0.3 cm}
 (e) $x'_+ < x \leq x_\mathrm{n}$:
the Carter-Penrose diagram [Fig. \ref{Fig-a3}.(e)]. As before [see
Fig \ref{Fig-a2}.(e)], it can be divided into left, middle and
right regions. The middle region consists of a timelike line
representing $r=+\infty$ (${\cal I}$) and two accelerated
Rindler-like horizon, $r_A^-=[A(x-x'_-)]^{-1}$ and
$r_A^+=[A(x-x'_+)]^{-1}$, that were already present in the $m=0$
and $q=0$ corresponding diagram [see Fig. \ref{Fig-a1}.(e)]. The
left and right regions both contain a timelike curvature
singularity ($r=0$). In addition, these left and right regions
contain, in the non-extremal case, an inner horizon ($r_-$) and an
outer horizon ($r_+$), in the extremal case they contain is a
single horizon ($r_+$), and in the naked case they have no horizon
(see however the physical interpretation of this case as a black
hole in the end of subsection \ref{sec:PI-mass}).

\subsection{\label{sec:PD B}
 Causal Structure of the $\bm{A=1/\ell}$ solutions}

The $A=1/\ell$ case was studied in detail in \cite{EHM1}. In
particular,  the causal structure of the massive uncharged
solution was discussed. For completeness, we will also present the
causal diagrams of the massless uncharged solution and of the
massive charged solution.

Once more, due to the lower restriction on the value of $y$
($-x\leq y$), the causal diagrams depend on the angular direction
$x$ we are looking at. We have to consider
 separately three distinct sets of angular directions (see
discussion on the text of Figs. \ref{g1}, \ref{g2} and \ref{g3}),
namely (a) $x_\mathrm{s}\leq x <0$, (b) $x =0$ and
 (c) $0 < x \leq x_\mathrm{n}$, where $x_\mathrm{s}=-1$ and
 $x_\mathrm{n}=+1$ when $m=0$ and $q=0$.
\subsubsection{\label{sec:PD B.1}
 \textbf{Massless uncharged solution ($\bm{m=0, q=0}$)}}

\begin{figure} [h]
\includegraphics*[height=4.5cm]{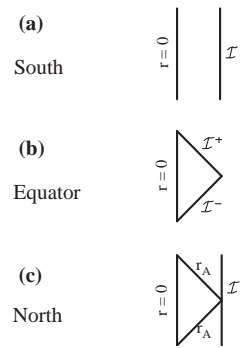}
   \caption{\label{Fig-b1}
Carter-Penrose diagrams of cases (a)-(c) discussed in the text of
section \ref{sec:PD B.1} concerning the $A=1/\ell\,,\,m=0\,$, and
$q=0$ C-metric.  $r_A=(A x)^{-1}$.  In diagrams (a) and (c) we
have to glue indefinitely copies of the represented figure in the
top and bottom regions of it.
 }
\end{figure}

In this case we have $x \in [x_\mathrm{s}=-1,x_\mathrm{n}=+1]$,
$x=\cos \theta$, ${\cal G}=1-x^2=\sin^2 \theta$, $\kappa=1$ and
${\cal F}(y) = y^2$ (see discussion on the text of Fig. \ref{g1}).
The angular surfaces $\Sigma$ ($t=$constant and $r=$constant) are
spheres free of conical singularities. The origin of the radial
coordinate $r$ has no curvature singularity and therefore both $r$
and $y$ can lie in the range $]-\infty,+\infty[$. However, in the
general case, where $m$ or $q$ are non-zero, there is a curvature
singularity at $r=0$. Since the discussion of the present section
is only a preliminary to that of the massive general case,
following \cite{AshtDray}, we will treat the origin $r=0$ as if it
had a curvature singularity and thus we admit that $r$ belongs to
the range $[0,+\infty[$ and $y$ lies in the region $-x\leq y <
+\infty$. The Carter-Penrose diagrams are drawn in Fig.
\ref{Fig-b1}. In case (c) $0 < x \leq x_\mathrm{n}$, and only in
this case, there is an accelerated horizon, $r_A=(Ax)^{-1}$.

\begin{figure} [h]
\includegraphics*[height=5.5cm]{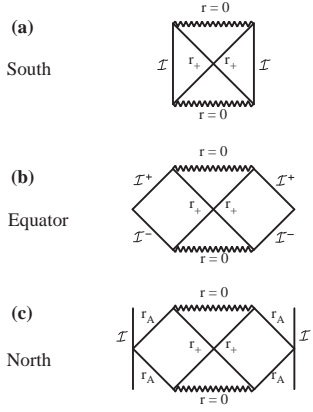}
   \caption{\label{Fig-b2}
 Carter-Penrose diagrams of cases (a)-(c) discussed in the text
of section \ref{sec:PD B.2} concerning the $A=1/\ell,
mA<3^{-3/2}$, and $q=0$ C-metric.  $r_A=(A x)^{-1}$ is a
degenerated horizon (see \cite{EHM1}).  In diagram (c) we have to
glue indefinitely copies of the represented figure in the top and
bottom regions of it.
 }
\end{figure}

\begin{figure}[t]
\includegraphics*[height=11.5cm]{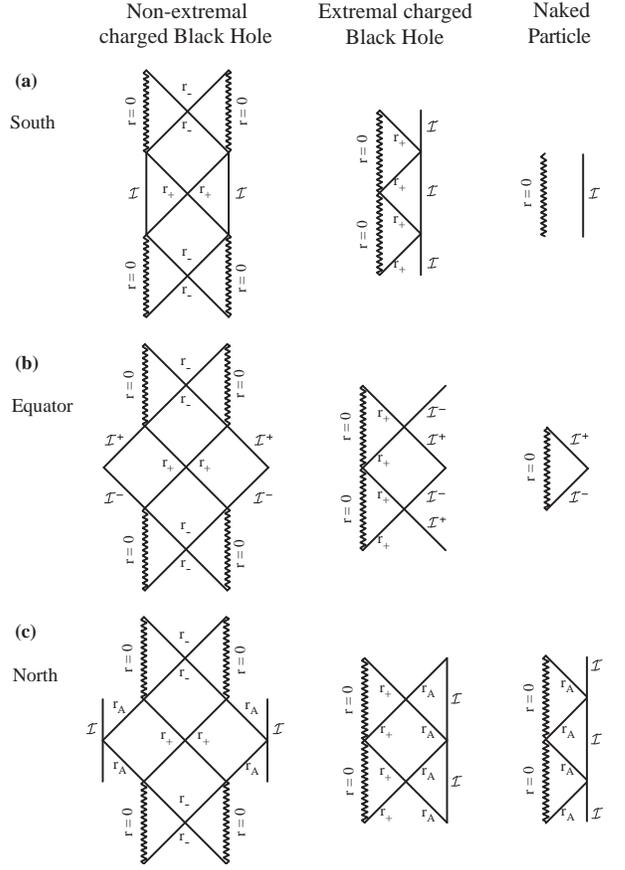}
   \caption{\label{Fig-b3}
Carter-Penrose diagrams of cases (a)-(c) discussed in the text of
section \ref{sec:PD B.3} concerning the charged massive C-metric
with $A=1/\ell$. The left column represents the non-extremal black
hole, the middle column represents the extremal black hole and the
right column represents the naked charged particle.  $r_A=(A
x)^{-1}$ is an accelerated horizon and $r_-$ and $r_+$ are charged
associated horizons.  In these diagrams  we have to glue
indefinitely copies of the represented figure in the top and
bottom regions of it.
 }
\end{figure}

\subsubsection{\label{sec:PD B.2}
 \textbf{Massive uncharged solution ($\bm{m >0}$, $\bm{q=0}$)}}

The causal diagrams of this solution were first presented in
\cite{EHM1} and are drawn in Fig. \ref{Fig-b2}. In the case (c) $0
< x \leq x_\mathrm{n}$, and only in this case, there is an
accelerated horizon, $r_A=(Ax)^{-1}$ which is degenerated (see
\cite{EHM1}). The Schwarzschild-like horizon is at
$r_+=A^{-1}[x+1/(2mA)]^{-1}$.

\subsubsection{\label{sec:PD B.3} \textbf{Massive charged
solution ($\bm{m >0}$, $\bm{q\neq0}$)}}

The Carter-Penrose diagrams of the solution for this range of
parameters is sketched in Fig. \ref{Fig-b3}. In these diagrams,
the left column represents the non-extremal black hole, the middle
column represents the extremal black hole and the right column
represents the naked charged particle. The row (a) describes the
solution seen from an angle that is between the south pole
(including) and the equator (excluding), row (b) applies only to
the equatorial direction, and row (c) describes the solution seen
from an angle between the equator (excluding) and the north pole
(including).

\subsection{\label{sec:PD C}
 Causal Structure of the $\bm{A<1/\ell}$ solutions}

The $A<1/\ell$ case was first analyzed in \cite{Pod}. We
complement it with the analysis of the causal structure.
Contrarily to the cases $A > 1/\ell$ and $A= 1/\ell$, the causal
diagrams of this spacetime do not depend on the angular direction
we are looking at. The reason for this feature is clearly
identified and explained in the discussion on the text of Figs.
\ref{g1}, \ref{g2} and \ref{g3}.

\subsubsection{\label{sec:PD C.1}
 \textbf{Massless uncharged solution ($\bm{m=0, q=0}$)}}

 The Carter-Penrose diagram is identical to the one of
the AdS solution ($A=0\,,\,m=0\,,\,q=0$) and is sketched in Fig.
\ref{Fig-b1}.(a). The origin has an acceleration $A$, as will be
seen in section \ref{sec:Phys_Interp}.

\subsubsection{\label{sec:PD C.2}
 \textbf{Massive uncharged solution ($\bm{m >0}$, $\bm{q=0}$)}}

The Carter-Penrose diagram is identical to the one of the
AdS-Schwarzschild solution ($A=0\,,\,m>0\,,\,q=0$) and is drawn in
Fig. \ref{Fig-b2}.(a). The origin has an acceleration $A$, as will
be seen in section \ref{sec:Phys_Interp}.

\subsubsection{\label{sec:PD C.3} \textbf{Massive charged
solution ($\bm{m >0}$, $\bm{q\neq0}$)}}

The Carter-Penrose diagrams  are identical to those of the
AdS$-$Reissner-Nordstr\"{o}m solution ($A=0\,,\,m>0\,,\,q\neq0$)
and is represented in Fig. \ref{Fig-b3}.(a). In this figure, the
non-extremal black hole is represented in the left column, the
extremal black hole is represented in the middle column, and the
naked charged particle is represented in the right column. The
origin has an acceleration $A$, as will be seen in section
\ref{sec:Phys_Interp}.

\section{\label{sec:Phys_Interp} PHYSICAL INTERPRETATION
OF THE A\lowercase{d}S C-metric}

The parameter $A$ that is found in the AdS C-metric is interpreted
as being an acceleration and the AdS C-metric with $A>1/\ell$
describes a pair of black holes accelerating away from each other
in an AdS background, while the AdS C-metric with $A\leq 1/\ell$
describes a single accelerated black hole. In this section we will
justify this statement.

In the Appendix it is shown that, when $A=0$, the general AdS
C-metric, Eq. (\ref{AdS C-metric}), reduces to the AdS
($m=0\,,\,q=0$), to the AdS-Schwarzschild ($m>0\,,\,q=0$), and to
the AdS$-$Reissner-Nordstr\"{o}m solutions ($m=0\,,\,q\neq0$).
Therefore, the parameters $m$ and $q$ are, respectively, the ADM
mass and ADM electromagnetic charge of the non-accelerated black
holes. Moreover, if we set the mass and charge parameters equal to
zero, even when $A\neq 0$, the Kretschmann scalar
 [see Eq. (\ref{R2})] reduces to the value expected for the AdS spacetime.
This indicates that the massless uncharged AdS C-metric is an AdS
spacetime in disguise.

\subsection{\label{sec:PI.2-BH}
$\bm{A>1/\ell}$. Pair of accelerated black holes}
In this section, we will first interpret case {\it 1. Massless
uncharged solution} ($m =0$, $q=0$), which is the simplest, and
then with the acquired knowledge we interpret cases {\it 2.
Massive uncharged solution} ($m>0$, $q=0$) and {\it 3. Massive
charged solution} ($m>0$, $q\neq0$). We will interpret the
solution following two complementary descriptions, the four
dimensional (4D) one and the five dimensional (5D).

\subsubsection{\label{sec:PI-4D} \textbf{The 4-Dimensional
description ($\bm{m =0}$, $\bm{q=0}$)}}

\begin{figure}[h]
\includegraphics*[height=1in]{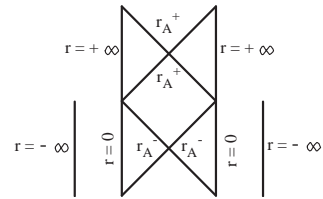}
   \caption{\label{Fig-e}
Extending the Carter-Penrose diagrams of Fig. \ref{Fig-a1} to
negative values of $r$, we obtain the AdS spacetime with its
origin being accelerated. $r_A^+ =[A(x-y_+)]^{-1}>0$ and
$r_A^-=[A(x+y_+)]^{-1}>0$. We have to glue indefinitely copies of
the represented figure in the top and bottom regions.
 }
\end{figure}

As we said in \ref{sec:PD A.1}, when $m=0$ and $q=0$ the origin of
the radial coordinate $r$ defined in Eq. (\ref{r}) has no
curvature singularity and therefore $r$ has the range
$]-\infty,+\infty[$. However, in the realistic general case, where
$m$ or $q$ are non-zero, there is a curvature singularity at $r=0$
and since the discussion of the massless uncharged solution was
only a preliminary to that of the massive general case, following
\cite{AshtDray}, we have treated the origin $r=0$ as if it had a
curvature singularity and thus we admitted that $r$ belongs to the
range $[0,+\infty[$. In these conditions we obtained the causal
diagrams of Fig. \ref{Fig-a1}. Note however that one can make a
further extension to include the negative values of $r$, enlarging
in this way the range accessible to the Kruskal coordinates $u'$
and $v'$. By doing this procedure we obtain the causal diagram of
the AdS spacetime. In Fig. \ref{Fig-e} we show the extension to
negative values of coordinate $r$ (so $-\infty <y<+\infty$) of the
Carter-Penrose diagrams of Fig. \ref{Fig-a1}. This diagram
indicates that the origin of the AdS spacetime, $r=0$, is
accelerating. The situation is analogous to the one that occurs in
the usual Rindler spacetime, $ds^2=-X^2dT^2+dX^2$. If one
restricts the coordinate $X$ to be positive one obtains an
accelerated origin that approaches a Rindler accelerated horizon.
However, by making an extension to negative values of $X$ one
obtains the Minkowski spacetime.

\begin{figure}[b]
\includegraphics*[height=1.1in]{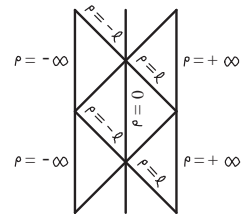}
   \caption{\label{Fig-d}
Carter-Penrose diagram of metric (\ref{metric-int}). We have to
glue indefinitely copies of the represented figure in the top and
bottom regions.
 }
\end{figure}

Now, we want to clearly identify the parameter $A$ that appears in
the AdS C-metric with the acceleration of its origin. To achieve
this aim, we recover the massless uncharged AdS C-metric defined
by Eq. (\ref{C-metric}) and Eq. (\ref{FG}) (with $m=0$ and $q=0$),
and after performing the following coordinate transformation
\begin{eqnarray}
& & \tau=\frac{\sqrt{\ell^2A^2-1}}{A} t \:,  \;\;\;\;\;
    \rho=\frac{\sqrt{\ell^2A^2-1}}{A} \frac{1}{y} \:, \nonumber \\
& & \theta = \arccos{x} \:,
     \;\;\;\;\; \phi = z \:,
  \label{transf-int}
  \end{eqnarray}
we can rewrite the massless uncharged AdS C-metric as
\begin{eqnarray}
 d s^2 = \frac{1}{\gamma^2}
 {\biggl [}-(1-\rho^2/\ell^2)d\tau^2+
 \frac{d\rho^2}{1-\rho^2/\ell^2} +\rho^2 d\Omega^2 {\biggl ]},
\label{metric-int}
 \end{eqnarray}
 with $d\Omega^2=d\theta^2+\sin^2\theta d \phi^2$ and
 \begin{eqnarray}
 \gamma=\sqrt{\ell^2A^2-1} + A\rho \cos\theta \:.
 \label{gamma}
 \end{eqnarray}
The causal diagram of this spacetime is drawn in Fig. \ref{Fig-d}.
Notice that the origin of the radial coordinate $\rho$ corresponds
to $y=+\infty$ and therefore to $r=0$, where $r$ has been
introduced in Eq. (\ref{r}). So, when we consider the massive AdS
C-metric there will be a curvature singularity at $\rho=0$ (see
section \ref{sec:CurvSing}).

To discover the meaning of parameter $A$ we consider the 4D
timelike worldlines described by an observer with $\rho=$constant,
$\theta=0$ and $\phi=0$ (see \cite{Lemos}). These are given by
$x^{\mu}(\lambda)=(\gamma \ell
\lambda/\sqrt{\ell^2-\rho^2},\rho,0,0)$, were $\lambda$ is the
proper time of the observer since the 4-velocity
$u^{\mu}=dx^{\mu}/d\lambda$ satisfies $u_{\mu}u^{\mu}=-1$. The
4-acceleration of these observers,
$a^{\mu}=(\nabla_{\nu}u^{\mu})u^{\nu}$, has a magnitude given by
\begin{eqnarray}
 |a_4|=\sqrt{a_{\mu}a^{\mu}}=\frac{\rho\sqrt{\ell^2A^2-1}+\ell^2A}
 {\ell\sqrt{\ell^2-\rho^2}}\:.
 \label{a}
 \end{eqnarray}
Since $a_{\mu}u^{\mu}=0$, the value $|a_4|$ is also the magnitude
of the 3-acceleration in the rest frame of the observer. From Eq.
(\ref{a}) we achieve the important conclusion that the origin of
the AdS C-metric, $\rho=0$ (or $r=0$), is being accelerated with a
constant acceleration whose value is precisely given by the
parameter $A$ that appears in the AdS C-metric. Moreover, at
radius $\rho=\ell$ [or $y=y_+$ defined in equation (\ref{F1})] the
acceleration is infinite which corresponds to the trajectory of a
null ray. Thus, observers held at $\rho=$constant see this null
ray as an acceleration horizon and they will never see events
beyond this null ray.

\subsubsection{\label{sec:PI-5D} \textbf{The 5-Dimensional
description ($\bm{m =0}$, $\bm{q=0}$)}}
In order to improve and clarify the physical aspects of the AdS
C-metric we turn now into the 5D representation of the solution.

The AdS spacetime can be represented as the 4-hyperboloid,
\begin{eqnarray}
-(z^0)^2+(z^1)^2+(z^2)^2+(z^3)^2-(z^4)^2=-\ell^2,
\label{hyperboloid}
 \end{eqnarray}
in the 5D Minkowski (with two timelike coordinates) embedding
spacetime,
\begin{eqnarray}
 d s^2 = -(dz^0)^2+(dz^1)^2+(dz^2)^2+(dz^3)^2-(dz^4)^2.
 \label{AdS}
 \end{eqnarray}
Now, the massless uncharged AdS C-metric is an AdS spacetime in
disguise and therefore our next task is to understand how the AdS
C-metric can be described in this 5D picture. To do this we first
recover the massless uncharged AdS C-metric described by Eq.
(\ref{metric-int}) and apply to it the coordinate transformation
\begin{eqnarray}
  \hspace{-0.3cm} & & \hspace{-0.3cm}
  z^0=\gamma^{-1}\sqrt{\ell^2-\rho^2}\,\sinh(\tau/\ell)\:,
  \;\;\;\;\; z^2=\gamma^{-1} \rho \sin\theta \cos\phi \:,
  \nonumber \\
  \hspace{-0.3cm} & & \hspace{-0.3cm}
  z^1=\gamma^{-1}\sqrt{\ell^2-\rho^2}\,\cosh(\tau/\ell)\:,
  \;\;\;\;\;  z^3=\gamma^{-1} \rho \sin\theta \sin\phi \:,
  \nonumber \\
  \hspace{-0.3cm} & & \hspace{-0.3cm}
  z^4=\gamma^{-1}[\sqrt{\ell^2A^2-1} \,\rho \cos\theta
  +\ell^2A]\:,
\label{AdS to AdS-c}
  \end{eqnarray}
where $\gamma$ is defined in Eq. (\ref{gamma}). Transformations
(\ref{AdS to AdS-c}) define an embedding of the massless uncharged
AdS C-metric into the 5D description of the AdS spacetime since
they satisfy Eq. (\ref{hyperboloid}) and take directly Eq.
(\ref{metric-int}) into Eq. (\ref{AdS}).
\begin{figure}[th]
\includegraphics*[height=2.6in]{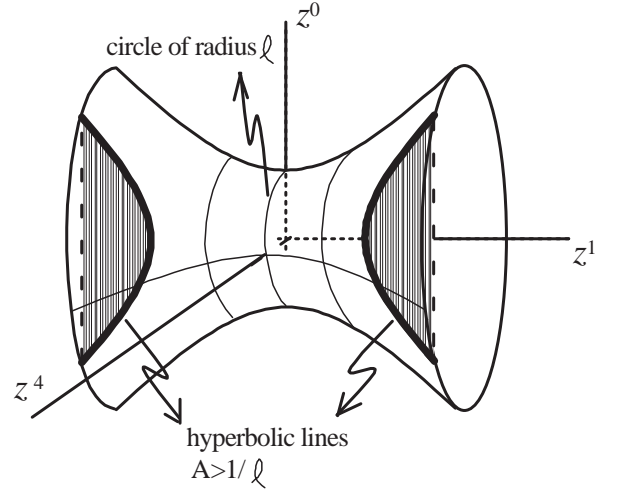}
\caption{\label{AdS-hyperb} AdS 4-hyperboloid embedded in the 5D
Minkowski spacetime with two timelike coordinates, $z^0$ and
$z^4$. The directions $z^2$ and $z^3$ are suppressed. The two
hyperbolic lines lying on the AdS hyperboloid result from the
intersection of the hyperboloid surface with the
$z^4$=constant$>\ell$ plane. They describe the motion of the
origin of the AdS C-metric with $A>1/\ell$.
 }
\end{figure}

\begin{figure}[t]
\includegraphics*[height=2.2in]{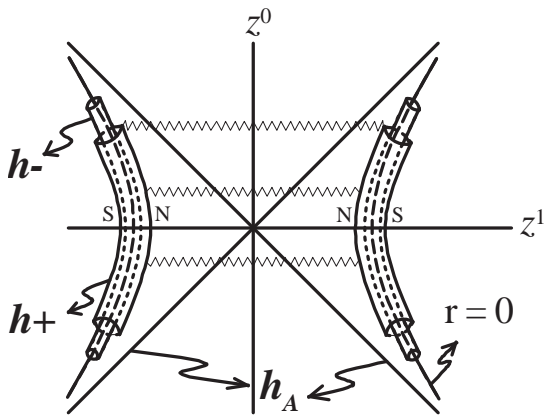}
\caption{\label{hyp} Schematic diagram representing the 5D
hyperbolic motion of two uniformly accelerating massive charged
black holes approaching asymptotically the Rindler-like
accelerated horizon ($h_A$). The inner and outer charged horizons
are represented by $h-$ and $h+$. The strut that connects the two
black holes is  represented by the zigzag lines. The north pole
direction is represented by $\rm{N}$ and the south pole direction
by $\rm{S}$.
 }
\end{figure}

 So, the massless uncharged AdS C-metric is an AdS spacetime, but
we can extract more information from this 5D analysis. Indeed, let
us analyze with some detail the properties of the origin of the
radial coordinate, $\rho=0$ (or $r=0$). This origin moves in the
5D Minkowski embedding spacetime according to [see Eq. (\ref{AdS
to AdS-c})]
\begin{eqnarray}
 & & z^2=0\;,\;\; z^3=0\;,\;\; z^4=\ell^2A /
 \sqrt{\ell^2A^2-1}\:>\ell
 \;\;\;\;\mathrm{and} \nonumber \\
 & & (z^1)^2-(z^0)^2=(A^2-1/\ell^2)^{-1}\equiv a_5^{-2} \:.
\label{rindler}
  \end{eqnarray}
These equations define two hyperbolic lines lying on the AdS
hyperboloid which result from the intersection of this hyperboloid
surface defined by Eq. (\ref{hyperboloid}) and the
$z^4$=constant$>\ell$ plane (see Fig. \ref{AdS-hyperb}). They tell
us that the origin is subjected to a uniform 5D acceleration,
$a_5$, and consequently moves along a hyperbolic worldline in the
5D embedding space, describing a Rindler-like motion (see Figs.
\ref{AdS-hyperb} and \ref{hyp}) that resembles the well-known
hyperbolic trajectory, $X^2-T^2=a^{-2}$, of an accelerated
observer in Minkowski space. But uniformly accelerated radial
worldlines in the 5D Minkowski embedding space are also uniformly
accelerated worldlines in the 4D AdS space \cite{DesLev}, with the
5D acceleration $a_5$ being related to the associated 4D
acceleration $a_4$ by $a_5^2=a_4^2-1/\ell^2$. Comparing this last
relation with Eq. (\ref{rindler}) we conclude that $a_4\equiv A$.
Therefore, and once again, we conclude that the origin of the AdS
C-metric is uniformly accelerating with a 4D acceleration whose
value is precisely given by the parameter $A$ that appears in the
AdS C-metric, Eq. (\ref{C-metric}), and this solution describes a
AdS space whose origin is not at rest as usual but is being
accelerated. Note that the origin of the usual AdS spacetime
describes the circle $(z^0)^2+(z^4)^2=\ell^2$ in the AdS
hyperboloid in contrast to the origin of the AdS C-metric with
$A>1/\ell$ whose motion is described by Eq. (\ref{rindler}). This
discussion allowed us to find the physical interpretation of
parameter $A$ and to justify its label. Notice also that the
original AdS C-metric coordinates introduced in Eq.
(\ref{C-metric}) cover only the half-space $z^1>-z^0$. The Kruskal
construction done in section \ref{sec:PD A} extended this solution
to include also the $z^1<-z^0$ region and so, in the extended
solution, $r=0$ is associated to two hyperbolas that represent two
accelerated points (see Fig. \ref{hyp}). These two hyperbolas
approach asymptotically the Rindler-like acceleration horizon
($r_A$), so called because it is is absent when $A=0$ and present
even when $A\neq 0$, $m=0$ and $q=0$.

\subsubsection{\label{sec:PI-mass}
\textbf{Pair of accelerated black holes ($\bm{m >0}$, $\bm{q \neq
0}$)}}
Now, we are in position to interpret the massive and charged
solutions that describe two black holes accelerating away from
each other. To see clearly this, let us look to the Carter-Penrose
diagrams near the equator, Fig. \ref{Fig-a1}.(c), Fig.
\ref{Fig-a2}.(c) and Fig. \ref{Fig-a3}.(c) (for the discussion
that follows we could, as well, look at the diagrams of case (d)
on these figures). Looking at these figures we can compare the
different features that belong to the massless uncharge case [Fig.
\ref{Fig-a1}.(c)], to the massive uncharged case [Fig.
\ref{Fig-a2}.(c)], and ending in the massive charged case [Fig.
\ref{Fig-a3}.(c)]. In Fig. \ref{Fig-a1}.(c) we identify the two
hyperbolas $r=0$ (represented by two timelike lines) approaching
asymptotically the Rindler-like acceleration horizon ($r_A$). When
we add a mass to the solution we conclude that each of these two
simple hyperbolas $r=0$ are replaced by the more complex structure
that represents a Schwarzschild black hole with its spacelike
curvature singularity and its horizon (these are represented by
$r_+$ in the left and right regions of Fig. \ref{Fig-a2}.(c)). So,
the two accelerating points $r=0$ have been replaced by two
Schwarzschild black holes that approach asymptotically the
Rindler-like acceleration horizon (represented by $r_A$ in the
middle region of Fig. \ref{Fig-a2}.(c)). The same interpretation
can be assigned to the massive charged solution. The two
hyperbolas $r=0$ of Fig. \ref{Fig-a1}.(c) are replaced by two
Reissner-Nordstr\"{o}m black holes (with its timelike curvature
singularity and its inner $r_-$ and outer $r_+$ horizons; see the
left and right regions of Fig. \ref{Fig-a3}.(c)) that approach
asymptotically the Rindler-like acceleration horizon already
present in the $m=0$ and $q=0$ causal diagram.

\vspace{0.1 cm}

The Carter-Penrose diagrams of cases (a) and (b) of Fig.
\ref{Fig-a2} and Fig. \ref{Fig-a3} indicate that an observer that
is looking through an angular direction which is in the vicinity
of the south pole does not see the acceleration horizon and
notices the presence of a single black hole. This is in agreement
with Fig. \ref{hyp}. Indeed, in this schematic figure, coordinates
$z^0$ and $z^1$ can be seen as Kruskal coordinates and we conclude
that an observer, initially located at infinity ($z^1=\infty$) and
moving inwards into the black hole along the south pole, passes
through the black hole horizons and hits eventually its curvature
singularity. Therefore, he never has the opportunity of getting in
contact with the acceleration horizon and with the second black
hole. This is no longer true for an observer that moves into the
black hole along an angular direction that is in the vicinity of
the north pole. In Fig. \ref{hyp} this observer would be between
the two black holes, at one of the points of the $z^0<0$ semi-axis
(say) and moving into the black hole. Clearly, this observer
passes through the acceleration horizon before crossing the black
hole horizons and hitting its curvature singularity. This
description agrees with cases (c), (d) and (e) of Fig.
\ref{Fig-a2} and Fig. \ref{Fig-a3} which describe the solution
along an angular direction which includes the equatorial plane
[case (c)] as well as the north pole [case (e)].

\vspace{0.1 cm}

The diagrams of the third column of Fig. \ref{Fig-a3} concerning
the naked case of the $A>1/\ell$ massive charged C-metric deserve
a comment. First, we stress that the term naked is employed in
this situation because the values of parameters $m$ and $q$ are
such that the solution has no charged associated horizons, i.e.,
in the notation used along this paper, $r_-$ and $r_+$ are not
present in these diagrams. However, these diagrams present an
interesting new feature. Indeed, looking at at rows (a) and (b) we
have a single accelerated naked particle, in rows (c)-(d) we find
two naked singularities approaching asymptotically the
acceleration horizon $r_A$ but in row (e) we have no longer two
naked singularities. More precisely, we have a kind of a single
AdS$-$Reissner-Nordstr\"{o}m  black hole with the curvature
singularity being provided by the mass and charge but with the
horizons having their origin in the acceleration and cosmological
constant.

\subsubsection{\label{sec:PI-strut} \textbf{Source of acceleration. The Strut}}

We can now ask what entity is causing the acceleration and where
it is localized. To achieve this aim, let us go back to the
massless uncharged AdS C-metric and consider radial worldlines
motions with $z^2=0$, $z^3=0$ and $z^4=$ constant or,
equivalently, with $\theta=0$, $\phi=$constant and
$\rho=$constant. These observers move along a Rindler-like
hyperbola described by [see Eq. (\ref{AdS to AdS-c})]
\begin{eqnarray}
(z^1)^2-(z^0)^2 &=& \frac{\ell^2-\rho^2}{(\sqrt{\ell^2A^2-1} +
A\rho)^2}  \:.
 \label{rindler.2}
  \end{eqnarray}
Since the right hand side of Eq. (\ref{rindler.2}) is smaller than
$a_5^{-2}$ defined in Eq. (\ref{rindler}), the north pole
$\theta_\mathrm{n}=0$ is localized between the hyperbolas
$(z^1)^2-(z^0)^2=a_5^{-2}$ in the $z^0,z^1$ diagram (see Fig.
\ref{hyp}). What does this means? When we put $m$ or $q$ different
from zero, each of the two hyperbolas assigned to $r=0$ represent
the accelerated motion of a black hole. Thus,
 Eq. (\ref{rindler.2}) tells us that the $\theta_\mathrm{n}=0$
axis points toward the other black hole, i.e., it is in the region
between the two black holes (see Fig. \ref{hyp}). The south pole
points along the symmetry axis towards spatial infinity. Now, in
section \ref{sec:ConSing}, we saw that parameter $\kappa$ has been
chosen in order to avoid a conical singularity at the south pole
[see Eq. (\ref{k-s})] and, by doing so, at the north pole is
localized a conical singularity. This is associated to a strut
that joins the two black holes and provides the acceleration of
the black holes. To confirm this, recall that either a straight
string or a strut have a metric described by
\cite{Vil,OscLem_string}
\begin{eqnarray}
 d s^2 = -dt^2+ dZ^2+ d\varrho^2+\varrho^2 d\tilde{\phi}^2,
\label{Vil-met}
 \end{eqnarray}
where $\tilde{\phi}=[1-\delta/(2\pi)]\phi$ and $0\leq\phi<2\pi$. A
string has $\delta>0$ and the geometry around it is conic, i.e.,
it is a plane with a deficit angle $\delta$, while a strut has
$\delta<0$. Their mass per unit length is $\mu=\delta/(8\pi)$ and
their interior energy-momentum tensor is
\begin{eqnarray}
T_{\alpha}^{\;\;\beta}=\mu\delta(X)\delta(Y)\mathrm{diag}(-1,0,0,-1),
\label{Vil-tens}
 \end{eqnarray}
where $X=\varrho\cos\tilde{\phi}$ and $Y=\varrho\sin\tilde{\phi}$
are the directions normal to the strut, and $\delta(X)$ and
$\delta(Y)$ are Dirac delta-functions. The pressure on the string
or in the strut satisfies $p=-\mu$. If $\mu>0$ we have a string,
if $\mu<0$ we have a strut. Now, turning to our case, the AdS
C-metric, Eq. (\ref{AdS C-metric}), near the north pole is given
by
\begin{eqnarray}
d s^2 &\sim& -r^2{\cal F}dt^2 + r^2{\cal F}^{-1}dy^2
                                                     \nonumber \\
& &+ {\biggr (}r^2d\theta^2 +
 \frac{\kappa^2}{4}{\biggl |}\frac{d G}{dx}{\biggl |}_{x_\mathrm{n}}
 r^2\theta^2 d\phi^2{\biggr )}\:,
 \label{N-metric}
 \end{eqnarray}
where $\kappa$ is defined in Eq. (\ref{k-s}) and the term between
the curved brackets is the induced metric in the plane normal to
the strut that connects the two black holes (along the $y$
direction) and will be labelled as $dX^2+dY^2$. The C-metric strut
has a mass per unit length given by
\begin{eqnarray}
\mu = \frac{1}{4}\frac{\delta_\mathrm{n}}{2\pi}=\frac{1}{4}
{\biggl (}1- {\biggl |}\frac{d {\cal G}}{dx}
 {\biggl |}^{-1}_{x_\mathrm{s}} {\biggl |}\frac{d {\cal G}}{dx}
 {\biggl |}_{x_\mathrm{n}}{\biggr )}  \:.
 \label{m.string}
  \end{eqnarray}
We have $|d_x {\cal G}|_{x_\mathrm{s}} < |d_x {\cal
G}|_{x_\mathrm{n}}$ and so $\mu$ is negative. To obtain the
pressure of the C-metric strut, we write Eq. (\ref{N-metric}) in a
Minkowski frame,
 $ds^2=-\theta^{(0)2}+\theta^{(1)2}+\theta^{(2)2}+\theta^{(3)2}$,
 with $\theta^{(A)}=e^{(A)}_{\;\;\;\;\alpha} dx^{\alpha}$ and
 $e^{(0)}_{\;\;\;\;0}=r\sqrt{{\cal F}}$, $e^{(1)}_{\;\;\;\;1}=r$,
 $e^{(2)}_{\;\;\;\;2}=r\theta k|d_x {\cal G}|_{x_\mathrm{n}}/2$ and
 $e^{(3)}_{\;\;\;\;3}=r/\sqrt{{\cal F}}$. In this Minkowski frame the
energy-momentum tensor, $T_{(A)}^{\;\;\;(B)}$, of the C-metric
strut is given by (\ref{Vil-tens}). In order to come back to the
coordinate basis frame and write the energy-momentum tensor of the
C-metric strut in this basis we use
 $T^{\alpha \beta}=e^{(A) \alpha}e_{(B)}^{\;\;\;\;\beta}T_{(A)}^{\;\;\;(B)}$
 and obtain
\begin{eqnarray}
 T^{\alpha \beta}=\mu (r^2{\cal F})^{-1}\delta(X)\delta(Y)
 \mathrm{diag}(1,0,0,-{\cal F}^2)\:.
\label{Vil-tens2}
 \end{eqnarray}
Defining the unit vector $\zeta=\partial/\partial y$ [so,
$\zeta^{\alpha}=(0,0,0,1)$], the pressure along the strut is
$T^{\alpha \beta}\zeta_{\alpha}\zeta_{\beta}$ and the pressure on
the C-metric strut is given by the integration over the $X$-$Y$
plane normal to the strut,
\begin{eqnarray}
p=\int dX dY \sqrt{^{(2)}g}\: T^{\alpha
\beta}\zeta_{\alpha}\zeta_{\beta}=-\mu\:.
 \label{Vil-pres}
 \end{eqnarray}
So, the pressure and mass density of the C-metric strut satisfy
the relation $p=-\mu$. Since $\mu$ is negative, at both ends of
the strut, one has a positive pressure pushing away the two black
holes.

Alternatively, instead of Eq. (\ref{k-s}), we could have chosen
for $\kappa$ the value $\kappa^{-1}=(1/2)|d_x {\cal
G}|_{x_\mathrm{n}}$. By doing so we would avoid the deficit angle
at the north pole ($\delta_\mathrm{n}=0$) and leave a conical
singularity at the south pole ($\delta_\mathrm{s}>0$). This option
would lead to the presence of a semi-infinite string extending
from each of the black holes towards infinity along the south pole
direction, which would furnish the acceleration. The mass density
of both strings is
 $\mu =(1/4)(1-|d_x {\cal G}|^{-1}_{x_\mathrm{n}}
 |d_x {\cal G}|_{x_\mathrm{s}})>0$
and the pressure on the string, $p=-\mu$, is negative which means
that each string is pulling the corresponding black hole towards
infinity.

\vspace{0.2 cm}

At this point, a remark is relevant. Israel and Khan \cite{bh_eq}
have found a solution that represents two (or more) collinear
Schwarzschild black holes interacting with each other in such a
way that allows dynamical equilibrium. In this solution, the two
black holes are connected by a strut that exerts an outward
pressure which cancels the inward gravitational attraction and so
the distance between the two black holes remains fixed
\cite{bh_eq}. The solution \cite{bh_eq} is valid for $\Lambda=0$
but, although it has not been done, it can be extended in
principle for generic $\Lambda$ and so the present remark holds
for generic $\Lambda$. Now, the C-metric solution reduces to a
single non-accelerated black hole free of struts or strings when
the acceleration parameter $A$ vanishes (see Appendix and section
\ref{sec:PI.1-BH.1}). Thus, when we take the limit $A=0$, the
C-metric does not reduce to the static solution of Israel and
Khan. The reason for this behavior can be found in the
Carter-Penrose diagrams of the C-metric. For example, looking into
Fig. \ref{Fig-a2}.(c) which represents the
 massive uncharged C-metric along the equator we conclude that a
null ray sent from the vicinity of one of the black holes can
 never cross the acceleration horizon ($r_A$) into the other black
 hole. So, if the two black holes cannot communicate through a
 null ray they cannot interact gravitationally. The only interaction
 that is present in the system is between the strut and each one
 of the black holes that suffer an acceleration
which is only furnished by the  strut's pressure. That the limit
$A=0$ does not yield the solution \cite{bh_eq} can also be
inferred from \cite{Yong}, where the C-metric is obtained from the
the two black hole solution of \cite{bh_eq} but through a singular
limit in which several quantities go appropriately to infinity.

\vspace{0.2 cm}

Ernst \cite{Ernst} has employed a Harrison-type transformation to
the $\Lambda=0$ charged C-metric in order to append a suitably
chosen external electromagnetic field. With this procedure the so
called Ernst solution is free of conical singularities at both
poles and the acceleration that drives away the two oppositely
charged Reissner-Nordstr\"{o}m black holes is totally provided by
the external electromagnetic field. In the AdS background we
cannot remove the conical singularities through the application of
the Harrison transformation \cite{Emparan}. Indeed, the Harrison
transformation does not leave invariant the cosmological term in
the action. Therefore, applying the Harrison transformation to
Eqs. (\ref{C-metric})-(\ref{potential}) does not yield a new
solution of the Einstein-Maxwell-AdS theory.

\subsubsection{\label{sec:PI-rad} \textbf{Radiative properties}}

The C-metric (either in the flat, de Sitter or anti-de Sitter
background) is an exact solution that is radiative. As noticed in
\cite{KW}, the gravitational radiation is present since the
complex scalar of the Newman-Penrose formalism,
$\Psi^4=-C_{\mu\nu\alpha\beta}
n^{\mu}\bar{m}^{\nu}n^{\alpha}\bar{m}^{\beta}$ (where
$C_{\mu\nu\alpha\beta}$ is the Weyl tensor and $\{l, n, m,
\bar{m}\}$ is the usual null tetrad of Newman-Penrose), contains a
term proportional to $r^{-1}$. Similarly, the charged version of
the C-metric includes, in addition, electromagnetic radiation. In
\cite{AshtDray}, it has been shown that the Bondi news functions
of the flat C-metric are indeed non-zero. These Bondi news
functions appear in the context of the Bondi method introduced to
study gravitational radiative systems. They are needed to
determine the evolution of the radiative gravitational field since
they carry the information concerning the changes of the system.
When at least one of them is not zero, the total Bondi mass of the
system decreases due to the emission of gravitational waves. The
Bondi news functions of the flat C-metric have been explicitly
calculated in \cite{Bic,PravPrav}. For a detailed review on the
radiative properties of the C-metric and other exact solutions see
\cite{PravPrav}. In AdS background these calculations have not
been carried yet. Indeed, AdS still lacks a peeling theorem.

\subsection{\label{sec:PI.1-BH}
$\bm{A= 1/\ell}$. Single accelerated black hole}

When $A= 1/\ell$ the AdS C-metric describes a single accelerated
black hole. The absence of a second black hole is clearly
indicated by the Carter-Penrose diagrams of Figs. \ref{Fig-b2} and
\ref{Fig-b3}.

This case has been studied in detail in \cite{EHM1} where the
Randall-Sundrum model in a lower dimensional scenario has
analyzed. In this scenario, the brane-world is a 2-brane moving in
a 4D asymptotically AdS background. They have shown that the AdS
C-metric with $A=1/\ell$ describes a black hole bound to the
Minkowski 2-brane. The brane tension is fine tuned relative to the
cosmological background acceleration and thus, $A=1/\ell$ is
precisely the acceleration that the black hole has to have in
order to comove with the 2-brane. They concluded that the AdS
C-metric describes the final state of gravitational collapse on
the brane-world.
 The causal structure of the massive uncharged solution (Fig.
\ref{Fig-b2}) has been first discussed in \cite{EHM1}. For
completeness, we have also presented the causal diagrams of the
massless uncharged solution in Fig. \ref{Fig-b1} and of the
non-extremal, extremal, and naked massive charged solutions in
Fig. \ref{Fig-b3}.

In \cite{EHM1} the coordinate transformation that takes the
massless uncharged AdS C-metric with $A= 1/\ell$ into the known
description of the AdS spacetime in Poincar\'e coordinates is
given. From there one can easily go to the 5D description on the
AdS hyperboloid. This 5D description can be also understood
directly from the limits on the solutions $A>1/\ell$ and
$A<1/\ell$ when $A \to 1/\ell$. Indeed, if we take the limit $A
\to 1/\ell$ in section \ref{sec:PI-5D} (where we have studied the
5D description of case $A>1/\ell$), one sees that the cut that
generates the two hyperbolic lines degenerates into two half
circles which, on identifying the ends of the AdS hyperboloid at
both infinities, yields one full circle. This means that the
trajectory of the origin of the AdS C-metric in the $A= 1/\ell$
case is a circle (which when one unwraps the hyperboloid to its
universal cover yields a straight accelerated line). As we will
see in the next subsection, for $A<1/\ell$ the trajectory of the
origin is a circle which, on taking the limit $A \to 1/\ell$,
still yields a circle. The two limits give the same result as
expected.
\subsection{\label{sec:PI.1-BH.1}
$\bm{A < 1/\ell}$. Single accelerated black hole}

 The $A<1/\ell$ case was
first analyzed in \cite{Pod}. We have complemented this work with
the analysis of the causal structure. The causal diagrams of this
spacetime are identical to the ones of the AdS ($m=0\,,\,q=0$)
[see Fig. \ref{Fig-b1}.(a)], of the AdS-Schwarzschild ($m
>0\,,\,q=0$)  [see Fig. \ref{Fig-b2}.(a)], and of the
AdS$-$Reissner-Nordstr\"{o}m solutions ($m >0\,,\,q\neq 0$) [see
Fig. \ref{Fig-b3}.(a)]. However, the curvature singularity of the
single black hole of the solution is not at rest but is being
accelerated, with the acceleration $A$ provided by an open string
that extends from the pole into asymptotic infinity.

\begin{figure}[t]
\includegraphics[height=0.8in]{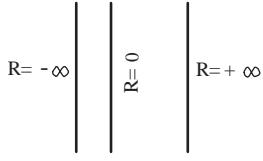}
   \caption{\label{Fig-f}
Carter-Penrose diagram of metric (\ref{metric-int2}). }
\end{figure}

As was done with the $A>1/\ell$ case, it is useful to interpret
the solution following two complementary descriptions, the 4D one
and the 5D. One first recovers the massless uncharged AdS C-metric
defined by Eq. (\ref{C-metric}) and Eq. (\ref{FG}) (with
$A<1/\ell$, $m=0$ and $q=0$), and after performing the following
coordinate transformation \cite{Pod}
\begin{eqnarray}
& & T=\frac{\sqrt{1-\ell^2A^2}}{A} t \:,  \;\;\;\;\;
    R=\frac{\sqrt{1-\ell^2A^2}}{A} \frac{1}{y} \:, \nonumber \\
& & \theta = \arccos{x} \:,
     \;\;\;\;\; \phi = z \:,
  \label{transf-int2}
  \end{eqnarray}
we can rewrite the massless uncharged AdS C-metric as
\begin{eqnarray}
 d s^2 = \frac{1}{\eta^2}
 {\biggl [}-(1+R^2/\ell^2)dT^2+
 \frac{dR^2}{1+R^2/\ell^2} +R^2 d\Omega^2 {\biggl ]},
 \nonumber \\
\label{metric-int2}
 \end{eqnarray}
with $\eta^{-1}=\sqrt{1-\ell^2A^2} + A R \cos\theta$ and
$d\Omega^2=d\theta^2+\sin^2\theta d \phi^2$. A procedure similar
to the one used to obtain (\ref{a}) indicates that an observer
describing 4D timelike worldlines  with $R=$constant, $\theta=0$
and $\phi=0$ suffers a 4-acceleration with magnitude given by
\begin{eqnarray}
 |a_4|=\frac{\ell^2A-R\sqrt{1-\ell^2A^2}}
 {\ell\sqrt{\ell^2+R^2}}\:.
 \label{a2}
 \end{eqnarray}
Therefore, the origin of the AdS C-metric, $R=0$, is being
accelerated with a constant acceleration whose value is precisely
given by $A$. The causal diagram of this spacetime is drawn in
Fig. \ref{Fig-f}. Notice that when we set $A=0$, Eq.
(\ref{metric-int2}) reduces to the usual AdS spacetime written in
static coordinates.
Now, to obtain the 5D description, one applies to Eq.
(\ref{metric-int2}) the coordinate
 transformation \cite{Pod},
\begin{eqnarray}
  \hspace{-0.3cm} & & \hspace{-0.3cm}
  z^0=\eta^{-1} \, \sqrt{\ell^2+R^2}\,\sin(T/\ell)\:,
  \;\;\;\;\;
  z^2=\eta^{-1} \, R \sin\theta \cos\phi \:,
  \nonumber \\
  \hspace{-0.3cm} & & \hspace{-0.3cm}
  z^4=\eta^{-1} \, \sqrt{\ell^2+R^2}\,\cos(T/\ell)\:,
  \;\;\;\;\;
  z^3=\eta^{-1} \, R \sin\theta \sin\phi \:,
  \nonumber \\
  \hspace{-0.3cm} & & \hspace{-0.3cm}
  z^1=\eta^{-1} \,[\sqrt{1-\ell^2A^2} \,R \cos\theta
  -\ell^2A]\:.
\label{AdS to AdS-c.2}
  \end{eqnarray}
Transformations (\ref{AdS to AdS-c.2}) define an embedding of the
massless uncharged AdS C-metric with $A<1/\ell$  into the 5D
description of the AdS spacetime since they satisfy Eq.
(\ref{hyperboloid}) and take directly Eq. (\ref{metric-int2}) into
Eq. (\ref{AdS}).

\begin{figure}
\includegraphics*[height=2.6in]{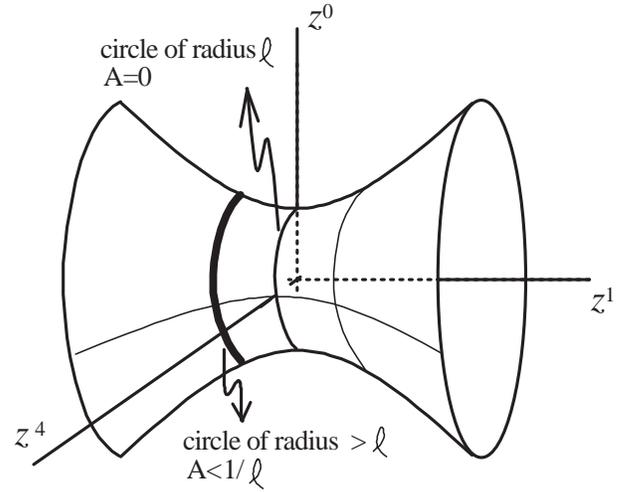}
\caption{\label{AdS-hyperb2} AdS 4-hyperboloid embedded in the 5D
Minkowski spacetime. The origin of the AdS C-metric with
$A<1/\ell$ moves in the hyperboloid along the circle with
$z^1$=constant$<0$. When $A=0$ this circle is at the plane $z^1=0$
and has a radius $\ell$.
 }
\end{figure}

The origin of the radial coordinate, $R=0$ moves in the 5D
Minkowski embedding spacetime according to [see Eq. (\ref{AdS to
AdS-c.2})]
\begin{eqnarray}
 & & z^1=-\ell^2A/\sqrt{1-\ell^2A^2}\;,\;\; z^2=0\;,\;\; z^3=0
 \;\;\;\;\mathrm{and} \nonumber \\
 & & (z^0)^2+(z^4)^2=(1/\ell^2-A^2)^{-1}\equiv a_5^{-2} \:.
\label{circ}
  \end{eqnarray}
So, contrarily to the case $A>1/\ell$ where the origin described a
Rindler-like hyperbolic trajectory [see Eq. (\ref{rindler})] that
suggests the presence of two black holes driving away from each
other in the extended  diagram, in the $A<1/\ell$ case the origin
describes a circle (a uniformly accelerated worldline) in the 5D
embedding space (see Fig. \ref {AdS-hyperb2}), indicating the
presence of a single trapped black hole in the AdS background.

To summarize and conclude, we present the global description on
the AdS hyperboloid of the AdS C-metric origin when the
acceleration $A$ varies from $+\infty$ to zero. When $A=+\infty$
the origin of the solution is represented in the hyperboloid by
two mutual perpendicular straight null lines at $45^{\rm o}$ that
result from the intersection of the hyperboloid surface defined by
Eq. (\ref{hyperboloid}) and the $z^4=\ell$ plane (see Figs.
\ref{AdS-hyperb} and \ref{hyp}). When $A$ belongs to
$]1/\ell,+\infty[$, the origin of the solution is represented by
two hyperbolic lines [Eq. (\ref{rindler})] lying on the AdS
hyperboloid and result from the intersection of Eq.
(\ref{hyperboloid}) and the $z^4$=constant$>\ell$ plane (see Fig.
\ref{AdS-hyperb}). As the acceleration approaches the value
$A=1/\ell$ the separation between the two hyperbolic lines
increases. When $A=1/\ell$ the separation between the two
hyperbolic lines becomes infinite and they collapse into two half
circles which, on identifying the ends of the AdS hyperboloid at
both infinities, yields one full circle in the $z^0-z^4$ plane at
infinite $z^1$. At this point we see again that the value
$A=1/\ell$ sets a transition stage between $A>1/\ell$ and
$A<1/\ell$. When $A$ belongs to $]0,1/\ell[$ the origin of the
solution is described again by a circle [Eq. (\ref{circ})] in the
$z^0-z^4$ plane but now at a constant $z^1<0$. As the acceleration
approaches the value $A=0$, the radius of this circle decreases
and when $A=0$ the circle has a radius with value $\ell$ and is at
$z^1=0$ (see Fig. \ref{AdS-hyperb2}) and we recover the usual AdS
solution whose origin is at rest.

\section{\label{sec:Conc}CONCLUSIONS}
The AdS C-metric found by Pleba\'nski and Demia\'nski
\cite{PlebDem} is characterized by a quite interesting new feature
when compared with the C-metric in flat or de Sitter backgrounds.
Indeed, contrarily to what happens in these two last solutions, in
the AdS background the solution only describes a pair of
accelerated black holes if the acceleration parameter satisfies
$A> 1/ \ell$, where $\ell$ is the cosmological length. The
acceleration is caused by a strut that connects the black holes.
The physical interpretation of the solutions has been essentially
taken from the analysis of the Carter-Penrose diagrams (following
the approach of Kinnersly and Walker \cite{KW} for the flat
C-metric), from the embedding of the massless uncharged solution
into the AdS 4-hyperboloid in a 5D Minkowski spacetime (with two
timelike coordinates), and from the physics of the strut. The
alternative approach of Bonnor \cite{Bonnor1} which puts the flat
C-metric into the Weyl form cannot be realized here, since the
introduction of the cosmological constant prevents such a
coordinate transformation.

For $A> 1/ \ell$, the embedding of the AdS C-metric into 5D
Minkowski space clearly shows that the origin of the AdS C-metric
solution is subjected to a uniform acceleration, and describes a
hyperbolic Rindler-like worldline in the AdS 4-hyperboloid
embedded in the 5D Minkowski space. To be more precise, the origin
is represented by two hyperbolic lines that approach
asymptotically the Rindler-like accelerated horizon, so called
because it is is absent when $A=0$ and present even when $A\neq
0$, $m=0$ and $q=0$. When we add a mass or a charge to the system
the causal diagrams indicate that now we have two
AdS-Schwarzschild or two AdS$-$Reissner-Nordstr\"{o}m black holes
approaching asymptotically the Rindler-like accelerated horizon.
We have proceeded to the localization of the conical singularity
present in the solution and concluded that it is between the two
black holes and along the symmetry axis (or alternatively from the
black holes out to infinity). When it is between the two black
holes, it is associated to a strut satisfying the relation
$p=-\mu>0$, where $p$ and $\mu$ are respectively the pressure on
the strut and its mass density. The pressure is positive, so it
points outwards into infinity and pulls the black holes apart,
furnishing their acceleration (as in the flat C-metric). When the
conical singularity points from each of the black holes into
infinity, it is associated to a  string with negative pressure
that pushes the black holes into infinity. From the analysis of
the Carter-Penrose diagrams we also concluded that the two black
holes cannot interact gravitationally. So, their acceleration is
provided only by the pressure exerted by the strut. This is the
reason why the limit $A=0$ of the C-metric does not reduce to the
static solution of Israel and Khan \cite{bh_eq}. This solution
describes two collinear Schwarzschild black holes connected by a
strut that exerts an outward pressure which cancels the inward
gravitational attraction and so the distance between the two black
holes remains fixed.

 For $A\leq 1/\ell$ the above procedure indicates the absence of a
 second black hole and so the solution describes a single black
 hole. In the AdS 4-hyperboloid, the origin of these solutions
 describes a circle in the plane defined by the two timelike
 coordinates. In a lower dimensional Randall-Sundrum
model, it has been shown that the $A = 1/\ell$ AdS C-metric
describes a black hole bound to a Minkowski 2-brane moving in a 4D
asymptotically AdS background \cite{EHM1}.

The C-metric solution for generic $\Lambda$ has been used
\cite{DGKT,HHR,Osc} to describe the final state of the quantum
process of pair creation of black holes, that once created
accelerate apart by an external field. In this context, we expect
that the black hole pair created in the AdS background must have
an acceleration
 $A>1/\ell$. Indeed, the AdS background is globally
contracting with an acceleration precisely equal to $1/\ell$.
Therefore,  a pair of virtual black holes created in this
background can only become real if the black hole acceleration is
greater than the contracting acceleration of the AdS background,
otherwise, the annihilation is inevitable. The quantum process
that might create the pair would be the gravitational analogue of
the Schwinger pair production of charged particles in an external
electromagnetic field. This would be one possible scenario to
create two exactly equal black holes with the same acceleration
that are described by the AdS C-metric solution with $A> 1/ \ell$.


\begin{acknowledgments}

This work was partially funded by Funda\c c\~ao para a Ci\^encia e
Tecnologia (FCT) through project CERN/FIS/43797/2001 and
PESO/PRO/2000/4014.  OJCD also acknowledges finantial support from the
portuguese FCT through PRAXIS XXI programme. JPSL thanks
Observat\'orio Nacional do Rio de Janeiro for hospitality.

\end{acknowledgments}

\appendix*
\section{\label{sec:Phys_Interp m,e}Mass and charge parameters}
In this Appendix, one gives the physical interpretation of
parameters $m$ and $q$ that appear in the AdS C-metric. We follow
\cite{Pod}.

Applying the coordinate transformations to  Eq. (\ref{C-metric})
(see \cite{Pod}),
\begin{eqnarray}
& & T=\sqrt{1-\ell^2A^2}A^{-1} t \:,  \;\;\;\;\;
    R=\sqrt{1-\ell^2A^2}(Ay)^{-1} \:, \nonumber \\
& & \theta = \int_{x}^{x_\mathrm{n}}{\cal{G}}^{-1/2}dx \:,
     \;\;\;\;\; \phi = z/\kappa \:,
  \label{mq}
  \end{eqnarray}
and setting $A=0$ (and $\kappa=1$) one obtains
 \begin{equation}
 d s^2 = - F(R)\, d T^2 +F^{-1}(R)\, d R^2 +R^2 (d \theta^2
  +\sin^2\theta\,d\phi^2) \:,
\label{mq2}
\end{equation}
where $F(R)=1+R^2/\ell^2 -2m/R + q^2/R^2$. So, when the
acceleration parameter vanishes, the AdS C-metric, Eq.
(\ref{C-metric}), reduces to the AdS-Schwarzschild and
AdS$-$Reissner-Nordstr\"{o}m black holes and the parameters $m$
and $q$ that are present in the AdS C-metric are precisely the ADM
mass and ADM electromagnetic charge of these non-accelerated black
holes. It should however be emphasized that the accelerated black
holes lose mass through radiative processes and so the
determination of the mass of the accelerated black holes would
require the calculation of the Bondi mass, which we do not here.


\end{document}